\documentclass[superscriptaddress,twocolumn,showpacs,a4paper,
amssymb,amsmath,nobibnotes,aps,prd,
showkeys,
nofootinbib,notitlepage]{revtex4-1}
\usepackage{verbatim}
\usepackage[T1]{fontenc}
\usepackage[utf8]{inputenc}
\usepackage[american]{babel}
\usepackage{epsfig}
\usepackage{graphicx,subcaption,caption}
\usepackage{booktabs}
\usepackage{multirow}
\usepackage{dcolumn}
\usepackage{amsmath}
\usepackage{mathtools}
\usepackage{amsfonts}
\usepackage{amssymb}
\usepackage{epstopdf}
\usepackage{bm}
\usepackage{siunitx}
\usepackage{braket}
\usepackage{enumitem}
\usepackage{soul}
\usepackage{todonotes}
\usepackage{rotating}

\usepackage{color}
\ExplSyntaxOn % providing \expandableinput
\cs_new:Npn \expandableinput #1
  { \use:c { @@input } { \file_full_name:n {#1} } }
\ExplSyntaxOff

\makeatletter % setting your search path
\providecommand*{\input@path}{}
\g@addto@macro\input@path{{"C:/Users/me/inpath/"}}% \input finds files in this directory
\makeatother

\definecolor{navyblue}{rgb}{0.0, 0.0, 0.5}
\definecolor{royalblue}{rgb}{0.25, 0.41, 0.88}
\definecolor{cadmiumgreen}{rgb}{0.0, 0.42, 0.24}
\definecolor{blue-violet}{rgb}{0.54, 0.17, 0.89}
\definecolor{darkviolet}{rgb}{0.58, 0.0, 0.83}
\definecolor{orange(colorwheel)}{rgb}{1.0, 0.5, 0.0}

\usepackage{hyperref}
\hypersetup{
    colorlinks=true, % false: boxed links; true: colored links
    linkcolor=royalblue, % color of internal links (change box color with linkbordercolor)
    citecolor=magenta}

\usepackage{booktabs}
\usepackage{multirow}
\usepackage{dcolumn}
\usepackage{colortbl}

\begin{document}

%better title to be found
%\title{Updated (model marginalized) constraints on cosmological parameters}
%\title{On the robustness of current cosmological parameter constraints}
\title{How robust are the parameter constraints extending the $\Lambda$CDM model?}

\date{\today}

\author{Stefano Gariazzo}
\email{stefano.gariazzo@unito.it}
\affiliation{Instituto de Fisica Teorica, CSIC-UAM
C/ Nicolás Cabrera 13-15, Campus de Cantoblanco UAM, 28049 Madrid, Spain}
\affiliation{Department of Physics, University of Turin, via P.\ Giuria 1, 10125 Turin (TO), Italy \looseness=-1}
\affiliation{Istituto Nazionale di Fisica Nucleare (INFN), Sezione di Torino, via P.\ Giuria 1, 10125 Turin (TO), Italy}

\author{William Giar\`e}
\email{w.giare@sheffield.ac.uk}
\affiliation{School of Mathematical and Physical Sciences, University of Sheffield, Hounsfield Road, Sheffield S3 7RH, United Kingdom}

\author{Olga Mena}
\email{omena@ific.uv.es}
\affiliation{Instituto de F{\'\i}sica Corpuscular  (CSIC-Universitat de Val{\`e}ncia), E-46980 Paterna, Spain}

\author{Eleonora Di Valentino}
\email{e.divalentino@sheffield.ac.uk}
\affiliation{School of Mathematical and Physical Sciences, University of Sheffield, Hounsfield Road, Sheffield S3 7RH, United Kingdom}

\preprint{}

\begin{abstract}
We present model-marginalized limits on the six standard $\Lambda$CDM cosmological parameters ($\Omega_{\rm c} h^2$, $\Omega_{\rm b} h^2$, $\theta_{\rm MC}$, $\tau_{\rm reio}$, $n_s$ and $A_s$), as well as on selected derived quantities ($H_0$, $\Omega_{\rm m}$, $\sigma_8$, $S_8$ and $r_{\rm drag}$), obtained by considering several extensions of the $\Lambda$CDM model and three independent Cosmic Microwave Background (CMB) experiments: the Planck satellite, the Atacama Cosmology Telescope, and South Pole Telescope. We also consider low redshift observations in the form of Baryon Acoustic Oscillation (BAO) data from the SDSS-IV eBOSS survey and Supernovae (SN) distance moduli measurements from the \textit{Pantheon-Plus} catalog. The marginalized errors are stable against the different minimal extensions of the $\Lambda$CDM model explored in this study. The largest impact on the parameter accuracy is produced by varying the effective number of relativistic degrees of freedom ($N_{\rm eff}$) or the lensing amplitude ($A_{\rm lens}$).
Nevertheless, the marginalized errors on some \textit{derived} parameters such as $H_0$ or $\Omega_{\rm m}$ can be up to two orders of magnitude larger than in the canonical $\Lambda$CDM scenario when considering only CMB data. In these cases, low redshift measurements are crucial for restoring the stability of the marginalized cosmological errors computed here.  Overall, our results underscore remarkable stability in the mean values and precision of the main cosmological parameters %making irrelevant the choice of different extensions of the basic $\Lambda$CDM model
once both high and low redshift probes are fully accounted for. %The very same results should be understood as a \textit{tool} to test exotic extensions of the $\Lambda$CDM model, as t
The marginalized values ~%should 
can be used in numerical analyses due to their robustness and slightly larger errors, providing a more realistic and conservative approach.
\end{abstract}

\maketitle

\section{Introduction}
\label{sec.Introduction}
The minimal $\Lambda$CDM model of cosmology has successfully explained a large number of cosmological observations at different scales. Within this minimal theoretical framework, the complexity of the Cosmic Microwave Background (CMB) acoustic peak structure or the matter power spectrum can be well described by only six fundamental parameters: the cold dark matter density ($\Omega_{\rm c} h^2$), the baryon density ($\Omega_{\rm b} h^2$), the angular size of the sound horizon at recombination ($\theta_{\rm MC}$), the optical depth ($\tau_{\rm reio}$), and the inflationary parameters such as the spectral index ($n_s$) and the amplitude ($A_s$) of scalar perturbations. Starting from these six parameters, a number of other important (derived) quantities can be computed, including the Hubble constant ($H_0$), the matter density ($\Omega_{\rm m}$), the clustering parameters ($\sigma_8$, and $S_8\equiv \sigma_8 \sqrt{\Omega_{\rm m}/0.3}$), or the sound horizon at the drag epoch $r_{\rm drag}$. 

Given the robustness and resilience shown in the results obtained over the years from different CMB experiments and large-scale structure probes, the mean values and errors of these parameters are often used to fit more exotic cosmological scenarios. For instance, the values of $\sigma_8$ and/or $r_{\rm drag}$ have been extensively used in the literature to test modified gravity models, interacting cosmologies, cosmologies with additional degrees of freedom, or non-standard neutrino scenarios. Nevertheless, it should be noted that the fitting values and uncertainties of these (primary and derived) quantities are obtained within some precise theoretical assumptions. In spite of the remarkable success and the simplicity of our best-working model of the universe, there is solid ground to believe that some missing physics phenomena are absent in $\Lambda$CDM. The most obvious example concerns neutrino masses: from neutrino oscillation experiments, we know that neutrinos are not massless; however, their total mass, required to be within the sub-eV regime, is totally unknown. While neutrinos should be massive, there could be additional missing ingredients in this minimal recipe of our universe. Just to mention a few concrete examples, dark energy could not be as simple as a cosmological constant (see, for instance, the indication at $3.9\sigma$ for Dynamical Dark Energy as obtained by the recent DESI release~\cite{DESI:2024mwx}) or there could be a small curvature component in our universe, as seems to be indicated by some analyses of the CMB measurements from the Planck 2018 legacy release with the baseline \texttt{Plik} likelihood~\cite{Planck:2019nip}, which seem to point towards the possibility of a closed Universe at more than three standard deviations~\cite{Aghanim:2018eyx,Handley:2019tkm,DiValentino:2019qzk,DiValentino:2020hov,Yang:2022kho}.

As a result, one straightforward question is how stable the mean values and errors of the main cosmological parameters (including the derived ones) describing the minimal $\Lambda$CDM scenario are. Answering this question is of primary importance because these values are often regarded as reference parameters when testing exotic cosmological scenarios. Therefore, a stable and robust estimation of these parameters constitutes a unique tool for cosmological analyses.

In this manuscript, we take a first step forward to address the issue raised above by computing the marginalized errors on the main cosmological parameters over a range of possible fiducial cosmologies based on minimal extensions of the $\Lambda$CDM model (see Refs.~\cite{Gariazzo:2018meg,diValentino:2022njd,DiValentino:2022edq,DiValentino:2023fei} for previous studies dealing with such a marginalization procedure). %Therefore, we not only provide an answer regarding the stability of our current (minimal) description of the universe within the $\Lambda$CDM framework, but also offer a set of robust mean values and errors that can be used, under suitable conditions as input when testing different non-standard physical scenarios.
Before proceeding, we stress that our analysis focuses on only minimal extensions to the cosmological model. On the one hand, we acknowledge that this could be perceived as a limitation of our results because we derive constraints marginalized over models with only one (or at most two) additional degrees of freedom typically fixed in $\Lambda$CDM. Consequently, we account only for extensions that do not significantly deviate from a baseline $\Lambda$CDM cosmology. However, it would also be impractical to attempt to account for all non-standard cosmological models that deviate more substantially from $\Lambda$CDM proposed over the past few years in the literature, and we see no compelling reason to selectively include only a few of them, as this could introduce selection biases and obscure our conclusions. In this sense, our approach is also advantageous because we test if, and to what extent, the results remain robust when relaxing some of the minimal but well-motivated assumptions within $\Lambda$CDM. As a result, we not only provide an assessment of the simplest assumptions of our current (minimal) description of the universe but also offer a set of reliable mean values and errors on key cosmological parameters that, under suitable conditions, can be regarded as more realistic and conservative, serving as inputs for testing a variety of non-standard physical scenarios.

The manuscript is organized as follows.  Sec.~\ref{sec.Methods} describes the methodology of the marginalization procedure, the data used in the analyses, and the different fiducial cosmologies included in the marginalization. Sec.~\ref{sec.Results} presents our results, which include tables and figures illustrating the deviation from the expected values within the different fiducial cosmologies. We conclude in Sec.~\ref{sec.Conclusions}.

\section{Methodology}
\label{sec.Methods}

\subsection{Basic statistics}
In the following, we shall review the basics of Bayesian statistics necessary for performing a marginalization over a number of different models.
We refer the reader to Refs.~\cite{Gariazzo:2018meg,diValentino:2022njd,DiValentino:2022edq,DiValentino:2023fei} for a more complete description of the full statistical analysis. 

Given a set of possible models $\mathcal{M}_i$ with prior probabilities $\pi_i$, we begin by computing their Bayesian evidences $Z_i$ for the selected dataset $d$. The posterior probability of model $i$ over all possible models, $p_i$, can be computed using:
\begin{equation}
\label{eq:modelposterior}
p_i = \frac{\pi_i Z_i}{\sum _j\pi_j Z_j}\,.
\end{equation}
If all models share a parameter or set of parameters $\theta$, we can use the model posterior probabilities $p_i$ together with the parameter posterior probability within model $i$, $p(\theta|d,\mathcal{M}_i)$, to compute the model-marginalized posterior probability $p(\theta|d)$ for $\theta$, given some data $d$:
\begin{equation}
\label{eq:mmposterior_def}
p(\theta|d) \equiv \sum_i p(\theta|d,\mathcal{M}_i) p_i \,.
\end{equation}
If all models have the same prior and using the Bayes factors $B_{i0}=Z_i/Z_0$ with respect to the favored model $\mathcal{M}_0$, the model-marginalized posterior becomes:
\begin{equation}
\label{eq:mmposterior}
p(\theta|d) = \frac{\sum_i p(\theta|d,\mathcal{M}_i) B_{i0} }{ \sum_j B_{j0} } \,.
\end{equation}
Notice that if the Bayes factors are large in favor of the preferred model (usually the simplest one), extensions of the minimal picture will not contribute significantly to the model-marginalized posterior.

In order to perform Bayesian model comparison using the Bayes factors and evaluate the strength of preference in favor of the best model, we follow a modified version of the Jeffreys' scale\footnote{Notice that our empirical scale, summarized in~Tab.\ref{tab:bayes}, deviates from the scale defined in the original Jeffreys' work~\cite{Jeffreys:1939xee}.} extracted from Ref.~\cite{Trotta:2008qt}, see Tab.~\ref{tab:bayes}.

\begin{table*}[htbp!]
\begin{center}

\begin{tabular}{c | c  | c  | c  }
\hline
\textbf{$|\ln B_0|$} 
& \textbf{Odds} 
& \textbf{Probability} 
& \textbf{Strength of evidence}  
\\
\hline
$<0.1$& $\lesssim 3:1$ & $<0.750$ & Inconclusive\\
$1$& $\sim 3:1$ & $0.750$ & Weak\\
$2.5$& $\sim 12:1$ & $0.923$ & Moderate\\
$5$& $\sim 150:1$ & $0.993$ & Strong\\
\hline\hline
\end{tabular}

\end{center}
\caption{Modified Jeffreys’ empirical scale to establish the strength of evidence when comparing two competing models.}
\label{tab:bayes}
\end{table*}

\begin{table}[htbp]
\begin{center}
\renewcommand{\arraystretch}{1.5}
\begin{tabular}{l@{\hspace{0. cm}}@{\hspace{1.5 cm}} c}
    \hline
    \textbf{Parameter}    & \textbf{Prior} \\
    \hline\hline
    $\Omega_{\rm b} h^2$         & $[0.005\,,\,0.1]$ \\
    $\Omega_{\rm c} h^2$         & $[0.001\,,\,0.99]$ \\
    $\log(10^{10}A_{\rm S})$     & $[2.91\,,\,3.91]$ \\
    $n_{\rm s}$                  & $[0.8\,,\, 1.2]$ \\
    $100\,\theta_{\rm {MC}}$     & $[0.5\,,\,10]$ \\
    $\tau$                       & $0.065 \pm 0.015$ \\
    \hline
    $\Omega_{\rm k} $     	     & $[-0.3\,,\,0.3]$\\
    $w_0$                        & $[-3\,,\,1]$ \\
    $w_a$                        & $[-3\,,\,2]$ \\
    $\alpha_{\rm s}$                  & $[-1\,,\, 1]$ \\
    $\sum m_{\nu}$ [eV]          & $[0.06\,,\,5]$\\
    $N_{\rm eff}$                & $[0.05\,,\,10]$\\
    $A_{\rm lens}$                & $[0\,,\,5]$\\
    \hline\hline
\end{tabular}
\caption{List of uniform prior distributions for cosmological parameters.}
\label{tab.Priors}
\end{center}
\end{table}

\subsection{Theoretical Models}

As pointed out in the introduction, a key point in our analysis is to derive robust bounds on the basic six $\Lambda$CDM  cosmological parameters and a number of derived ones ($H_0$, $\Omega_{\rm m}$, $\sigma_8$, $S_8$ and $r_{\rm {drag}}$) by marginalizing over a plethora of possible background cosmologies,
obtained by extending the basic 6-parameters $\Lambda$CDM scenario. We recall here that $S_8\equiv\sigma_8 \sqrt{\Omega_{\rm m}/0.3}$ and $r_{\rm drag}$ is the sound horizon at the baryon drag epoch, the comoving distance a wave can travel prior to $z_{\rm drag}$, when baryons and photons decouple.

Therefore, along with the six $\Lambda$CDM parameters, we also include  several extensions of this minimal model, enlarging the parameter space including one or more degrees of freedom, such as
a running of the scalar spectral index ($\alpha_s$), 
a curvature component ($\Omega_k$),
the dark energy equation of state -- either parameterized via one single parameter ($w_0$), or via two parameters ($w_0$ and $w_a$) -- the total neutrino mass ($\sum m_\nu$), the effective number of relativistic degrees of freedom ($N_{\rm eff}$), and the lensing amplitude ($A_{\rm lens}$). 
A short description of each parameter follows.
\begin{itemize}
\item The running of scalar spectral index, $\alpha_s$. In simple inflationary models, the running of the spectral index is typically very small. However, specific models can produce a large running over a range of scales accessible to CMB experiments. Indeed, a non-zero value of $\alpha_s$ alleviates the $\sim 2.7\sigma$ discrepancy in the value of the scalar spectral index $n_{\rm s}$ measured by \emph{Planck} ($n_{\rm s}=0.9649\pm 0.0044$)~\cite{Planck:2018vyg} and by the \emph{Atacama Cosmology Telescope}  (ACT) ($n_{\rm s}=1.008\pm 0.015$)~\cite{ACT:2020gnv}, see Refs.~\cite{DiValentino:2022rdg, DiValentino:2022oon,Giare:2022rvg}. ACT and Planck are actually in tension regarding the estimate of $\alpha_s$ (see Ref.~\cite{Forconi:2021que}).

\item Curvature density, $\Omega_k$. Recent data analyses of the CMB temperature and polarization spectra from Planck 2018 team exploiting the baseline \emph{Plik} likelihood suggest that our Universe could have a closed geometry at more than three standard deviations~\cite{Planck:2018vyg,Handley:2019tkm,DiValentino:2019qzk,Semenaite:2022unt}. These hints mostly arise from TT observations, that would otherwise show a lensing excess~\cite{DiValentino:2020hov,Calabrese:2008rt,DiValentino:2019dzu}. 
In addition, analyses exploiting the \emph{CamSpec} TT likelihood~\cite{Efstathiou:2019mdh,Rosenberg:2022sdy} point to a closed geometry of the Universe with a significance above 99\% CL. However, this indication is reduced with the new \emph{HiLLiPoP} likelihood~\cite{Tristram:2023haj}.
Furthermore, an indication for a closed universe is also present in the BAO data, using Effective Field Theories of Large Scale Structure~\cite{Glanville:2022xes}. These recent findings strongly motivate to leave the curvature of the Universe as a free parameter~\cite{Anselmi:2022uvj} and obtain marginalized limits on the different cosmological parameters, accounting also for this context. 
    
\item Dark Energy equation of state, $w_0$ or $w_0 w_a$. 
Cosmological bounds become weaker if the dark energy equation of state is taken as a variable quantity. Indeed, the dark energy equation of state, not that associated to a cosmological constant $\Lambda$, has non-trivial degeneracies with a number of cosmological parameters, such as the Hubble constant or the total matter density. Even if current data fits well with the assumption of a cosmological constant within the minimal $\Lambda$CDM scenario\footnote{See also recent Ref.~\cite{Escamilla:2023oce} for a review on the constraints on the Dark Energy equation of state resulting from different cosmological and astrophysical data.}
(except in the case of DESI observations~\cite{DESI:2024mwx}) the question of having an equation of state parameter different from $ -1 $ remains certainly open. Along with constant dark energy equation of state models, in this paper we also consider the possibility of having a time-varying $ w(a) $ described by the Chevalier-Polarski-Linder parametrizazion (CPL)~\cite{Chevallier:2000qy,Linder:2002et}:
\begin{equation}
\label{eq:cpl}
    w(a) = w_0 + (1-a)w_a
\end{equation}
where $ a $ is the scale factor and is $ a_0 = 1 $ at the present time, $ w(a_0)=w_0 $ is the value of the equation of state parameter today. Dark energy changes the distance to the CMB consequently pushing it further (closer) if $w < -1$ ($w > -1$) from us. This effect can be balanced, for instance, by having a different matter density or a shifted value of $H_0$.

\item Neutrino mass, $\sum m_\nu$. Current cosmological data from the Planck CMB  satellite, the SDSS-III and SDSS-IV galaxy clustering surveys~\cite{Dawson:2015wdb,Alam:2020sor} and the Pantheon Supernova Ia provides the most constraining neutrino mass bound to date, $\sum m_\nu<0.09$~eV at $95\%$~CL~\cite{DiValentino:2021hoh}, mostly due to Redshift Space Distortions analyses from the SDSS-IV eBOSS survey (see also Ref.~\cite{Palanque-Delabrouille:2019iyz} for a very competitive limit), implying that \emph{six million neutrinos cannot weigh more than one electron.} More recently, the DESI collaboration reported a even stronger limit $\sum m_\nu<0.072$ (0.113)~eV at 95\% CL if the prior on the sum of the neutrino masses is assumed to be $\sum m_\nu \geq0$ (0.059)~eV \cite{DESI:2024mwx}. A larger amount of neutrino masses will shift the Hubble parameter towards smaller values. The value of the current matter energy density will be larger. This will also have a non-negligible impact in the value of the clustering parameter $\sigma_8$. 
    
\item The effective number of relativistic degrees of freedom, $N_{\rm eff}$. Our current knowledge of the relativistic degrees of freedom at decoupling demonstrates that  $N_{\rm eff}$ is close to 3 as measured by CMB observations ($N_{\rm eff}=2.99^{+0.34}_{-0.33}$ at 95\% confidence level (CL) \cite{Planck:2018vyg}) or BBN abundances (e.g.~$N_{\rm eff}=2.87^{+0.24}_{-0.21}$ at 68\% CL \cite{Consiglio:2017pot}) independently.  A larger value of $N_{\rm eff}$ will imply more radiation in the early universe and will be degenerate with the matter density, the amplitude of the primordial power spectrum and the Hubble constant.

\item The lensing amplitude, $A_{\rm lens}$. CMB anisotropies get blurred due to gravitational lensing by the large scale structures of the Universe: photons from different directions are mixed and the peaks at large multipoles are smoothed. The amount of lensing within a given cosmology can be changed by means of the factor $A_{\rm{lens}}$~\cite{Calabrese:2008rt}, the so-called lensing amplitude and, \emph{a priori}, an unphysical parameter. Within the minimal $\Lambda$CDM scenario, $A_{\rm{lens}}=1$. Planck measurements of the CMB temperature and polarization indicate that data shows a preference for additional lensing, suggesting $A_\mathrm{lens} > 1$ at 3$\sigma$. CMB lensing can be also extracted from CMB observations via a four-point correlation function. If this independent measurement is added the tension is ameliorated, albeit it is still above the canonical one by about $2\sigma$. This lensing anomaly could have its origin in other physics effects unrelated to lensing. Also in this case, this indication is reduced with the new \emph{HiLLiPoP} likelihood~\cite{Tristram:2023haj}. The lensing amplitude also shows non-negligible degeneracies with a number of cosmological parameters, as for instance, with the dark matter mass-energy density or with the reionization optical depth $\tau_{\rm reio}$~\cite{Giare:2023ejv}.
    
\end{itemize}

\subsection{Statistical Analyses and Likelihoods}
Our statistical analysis of CMB and large scale structure probes
is based on the public code \texttt{COBAYA}~\citep{Torrado:2020xyz},
of which we make use of the Monte Carlo Markov Chain (MCMC) sampler,
originally developed for \texttt{CosmoMC}~\cite{Lewis:2002ah}.
The sampler allows to perform parameter space exploration with speed hierarchy implementing the ``fast dragging'' procedure developed in~\cite{Neal:2005}. The prior distributions for the parameters involved in our analysis are chosen to be uniform along the range of variation (see Tab.~\ref{tab.Priors}) with the exception of the optical depth for which the prior distribution is chosen accordingly to the CMB datasets as discussed below. 
From the MCMC results, we compute Bayesian evidences thanks to the publicly available package \texttt{MCEvidence},\footnote{\href{https://github.com/yabebalFantaye/MCEvidence}{github.com/yabebalFantaye/MCEvidence}~\cite{Heavens:2017hkr,Heavens:2017afc}.} properly modified to be compatible with \texttt{COBAYA}.
It has been shown in the past, see e.g.~\cite{DiValentino:2022edq,DiValentino:2023fei},
that the \texttt{MCEvidence} algorithm can accurately reproduce the Bayes factors obtained with nested sampling tools such as \texttt{PolyChord}~\cite{Handley:2015fda,Handley:2015aa},
but with shorter computation time.
A reasonable estimate of the numerical uncertainties on the Bayes factors obtained here is therefore $\sigma(\log B)\sim0.5$ \cite{DiValentino:2022edq}.

Concerning the cosmological and astrophysical observations, our baseline data-sets and likelihoods include:

\begin{itemize}[leftmargin=*]

    \item Planck 2018 temperature and polarization (plik TT TE EE) likelihoods, which also include low multipole data ($\ell < 30$)~\cite{Aghanim:2019ame,Aghanim:2018eyx,Akrami:2018vks}, in combination with the Planck 2018 lensing likelihood~\cite{Aghanim:2018oex}, reconstructed from measurements of the power spectrum of the lensing potential. This dataset is referred to as \textit{\textbf{Planck}}.
	
    \item Atacama Cosmology Telescope temperature and polarization anisotropy DR4 likelihood, in combination with the gravitational lensing DR6 likelihood covering 9400 deg$^2$ reconstructed from CMB measurements made by the Atacama Cosmology Telescope from 2017 to 2021~\cite{ACT:2023kun,ACT:2023dou}. In our analysis for the lensing spectrum we include only the conservative range of lensing multipoles $40 < \ell < 763$.  We consider a Gaussian prior on $\tau = 0.065 \pm 0.015$, as prescribed in~\cite{ACT:2020gnv}. We refer to this dataset as \textit{\textbf{ACT}}.

    \item South Pole Telescope temperature and polarization (TT TE EE) likelihood~\cite{SPT-3G:2022hvq}. Also in this case we consider a Gaussian prior on $\tau = 0.065 \pm 0.015$, We refer to this dataset as \textit{\textbf{SPT}}.
    
    \item Local Universe observations in the form of\\
    \\
    i) Baryon Acoustic Oscillation data from the finalized SDSS-IV eBOSS survey. These data encompass both isotropic and anisotropic distance and expansion rate measurements, as outlined in Table 3 of Reference~\cite{eBOSS:2020yzd}.
    \\
    \\
    ii) Distance modulus measurements of Type Ia supernovae obtained from the \textit{Pantheon-Plus} sample~\cite{Brout:2022vxf}. This dataset comprises 1701 light curves representing 1550 unique Type Ia supernovae, spanning a redshift range from 0.001 to 2.26. \\
    \\
    We will refer to the combination of these two likelihoods as \textbf{\textit{low-z}}.    

\end{itemize}

\section{Results}
\label{sec.Results}

\begin{figure*}[htpb!]
\centering
\includegraphics[width=0.7\textwidth]{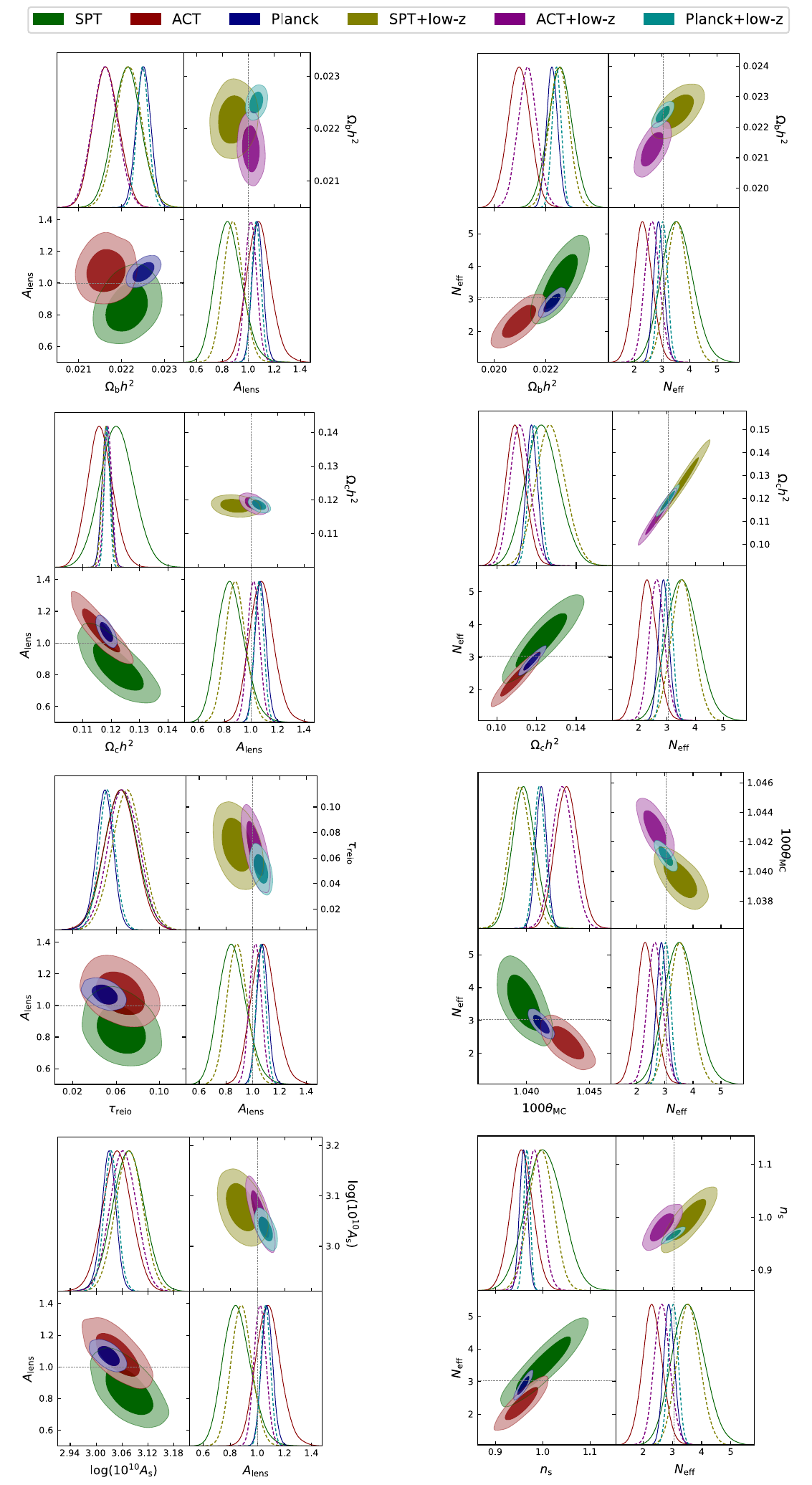}
\caption{Marginalized 1D and 2D posteriors involving the six standard $\Lambda$CDM parameters and other beyond-$\Lambda$CDM parameters that exhibit the most significant correlations and degeneracy lines with the former, across all the different CMB and CMB+low-$z$ datasets analyzed in this work.}
\label{fig:corr_main_6}
\end{figure*}

\begin{table*}[htpb!]
\centering
\renewcommand{\arraystretch}{1.2}
\resizebox{0.9\textwidth}{!}{\begin{tabular}{l ccc ccc}
\toprule
\textbf{Parameter} & \textbf{Planck} & \textbf{ACT} & \textbf{SPT} & \textbf{Planck+low-$z$} & \textbf{ACT+low-$z$} & \textbf{SPT+low-$z$} \\
\midrule
$\Omega_\mathrm{b} h^2$ & $0.02240_{-0.00016}^{+0.00015}$ & $0.02147_{-0.00043}^{+0.00039}$ & $0.02220_{-0.00032}^{+0.00033}$ & $0.02242_{-0.00013}^{+0.00015}$ & $0.02147\pm0.00034$ & $0.02223\pm0.00032$ \\
 & $(0.02240\pm0.00031)$ & $(0.02147_{-0.0010}^{+0.00071})$ & $(0.02220_{-0.00064}^{+0.00066})$ & $(0.02242_{-0.00026}^{+0.00028})$ & $(0.02147_{-0.00068}^{+0.00065})$ & $(0.02223_{-0.00061}^{+0.00066})$ \\
\midrule
$\Omega_\mathrm{c} h^2$ & $0.1197_{-0.0015}^{+0.0013}$ & $0.1187_{-0.0062}^{+0.0040}$ & $0.1171_{-0.0048}^{+0.0059}$ & $0.11925_{-0.00098}^{+0.0010}$ & $0.1189\pm0.0017$ & $0.1178_{-0.0023}^{+0.0024}$ \\
 & $(0.1197_{-0.0031}^{+0.0026})$ & $(0.1187_{-0.016}^{+0.0059})$ & $(0.1171_{-0.0089}^{+0.013})$ & $(0.1192\pm0.0025)$ & $(0.1189_{-0.010}^{+0.0033})$ & $(0.1178_{-0.0041}^{+0.016})$ \\
\midrule
$100\theta_\mathrm{MC}$ & $1.04095\pm0.00032$ & $1.04227_{-0.00080}^{+0.00084}$ & $1.04009_{-0.00079}^{+0.00081}$ & $1.04101_{-0.00030}^{+0.00029}$ & $1.04219\pm0.00065$ & $1.04019\pm0.00070$ \\
 & $(1.04095_{-0.00062}^{+0.00064})$ & $(1.0423_{-0.0015}^{+0.0019})$ & $(1.0401\pm0.0016)$ & $(1.04101_{-0.00059}^{+0.00058})$ & $(1.0422_{-0.0013}^{+0.0014})$ & $(1.0402_{-0.0015}^{+0.0014})$ \\
\midrule
$\tau_\mathrm{reio}$ & $0.0529_{-0.0073}^{+0.0080}$ & $0.064\pm0.015$ & $0.061\pm0.015$ & $0.0565_{-0.0069}^{+0.0075}$ & $0.066_{-0.011}^{+0.013}$ & $0.063_{-0.013}^{+0.014}$ \\
 & $(0.053_{-0.015}^{+0.016})$ & $(0.064\pm0.029)$ & $(0.061\pm0.029)$ & $(0.057_{-0.014}^{+0.015})$ & $(0.066_{-0.023}^{+0.025})$ & $(0.063_{-0.026}^{+0.027})$ \\
\midrule
$n_\mathrm{s}$ & $0.9658_{-0.0045}^{+0.0044}$ & $0.990_{-0.031}^{+0.017}$ & $0.966_{-0.022}^{+0.021}$ & $0.9670_{-0.0040}^{+0.0036}$ & $0.990_{-0.023}^{+0.014}$ & $0.969_{-0.017}^{+0.020}$ \\
 & $(0.9658_{-0.0089}^{+0.0093})$ & $(0.990_{-0.062}^{+0.033})$ & $(0.966_{-0.047}^{+0.048})$ & $(0.9670_{-0.0079}^{+0.0075})$ & $(0.990_{-0.039}^{+0.027})$ & $(0.969_{-0.033}^{+0.053})$ \\
\midrule
$\log(10^{10} A_\mathrm{s})$ & $3.041\pm0.016$ & $3.050_{-0.032}^{+0.034}$ & $3.054_{-0.032}^{+0.035}$ & $3.048_{-0.014}^{+0.015}$ & $3.061_{-0.025}^{+0.024}$ & $3.061\pm0.029$ \\
 & $(3.041_{-0.033}^{+0.032})$ & $(3.050_{-0.069}^{+0.063})$ & $(3.054_{-0.064}^{+0.071})$ & $(3.048_{-0.029}^{+0.030})$ & $(3.061_{-0.049}^{+0.045})$ & $(3.061_{-0.058}^{+0.057})$ \\
\midrule
$H_0$ [km/s/Mpc] & $67.3_{-5.8}^{+33}$ & $67.2_{-10}^{+6.1}$ & $68.3_{-4.0}^{+19}$ & $67.64_{-0.41}^{+0.49}$ & $67.33_{-0.60}^{+0.66}$ & $67.84_{-0.78}^{+0.67}$ \\
 & $(67.3_{-6.9}^{+33})$ & $(67_{-10}^{+33})$ & $(68_{-18}^{+28})$ & $(67.64_{-0.95}^{+1.1})$ & $(67.3_{-2.9}^{+1.3})$ & $(67.8_{-1.5}^{+4.0})$ \\
\midrule
$\Omega_{\mathrm{m}}$ & $0.315_{-0.016}^{+0.052}$ & $0.313_{-0.024}^{+0.048}$ & $0.296_{-0.14}^{+0.093}$ & $0.3104_{-0.0051}^{+0.0059}$ & $0.3116_{-0.0072}^{+0.0061}$ & $0.3056_{-0.0082}^{+0.0094}$ \\
 & $(0.315_{-0.17}^{+0.077})$ & $(0.31_{-0.17}^{+0.11})$ & $(0.30_{-0.15}^{+0.33})$ & $(0.310\pm0.011)$ & $(0.312\pm0.013)$ & $(0.306_{-0.013}^{+0.015})$ \\
\midrule
$\sigma_8$ & $0.810_{-0.026}^{+0.24}$ & $0.830_{-0.074}^{+0.033}$ & $0.804_{-0.21}^{+0.067}$ & $0.811_{-0.015}^{+0.014}$ & $0.8322_{-0.0095}^{+0.0088}$ & $0.811\pm0.014$ \\
 & $(0.810_{-0.034}^{+0.25})$ & $(0.830_{-0.095}^{+0.25})$ & $(0.80_{-0.28}^{+0.12})$ & $(0.811_{-0.015}^{+0.014})$ & $(0.832_{-0.022}^{+0.018})$ & $(0.811_{-0.027}^{+0.034})$ \\
\midrule
$S_8$ & $0.832_{-0.025}^{+0.030}$ & $0.848\pm0.025$ & $0.789_{-0.073}^{+0.063}$ & $0.827_{-0.012}^{+0.011}$ & $0.847\pm0.013$ & $0.821\pm0.019$ \\
 & $(0.832_{-0.096}^{+0.049})$ & $(0.848_{-0.086}^{+0.051})$ & $(0.79_{-0.16}^{+0.13})$ & $(0.827_{-0.023}^{+0.021})$ & $(0.847_{-0.026}^{+0.025})$ & $(0.821_{-0.037}^{+0.040})$ \\
\midrule
$r_{\rm drag}$ [Mpc] & $147.16\pm0.31$ & $148.38_{-0.98}^{+1.1}$ & $147.9_{-1.7}^{+1.2}$ & $147.24_{-0.32}^{+0.30}$ & $148.38_{-0.66}^{+0.69}$ & $147.85_{-0.72}^{+0.68}$ \\
 & $(147.16_{-0.78}^{+0.85})$ & $(148.4_{-1.6}^{+3.6})$ & $(147.9_{-3.3}^{+2.3})$ & $(147.2_{-1.9}^{+2.1})$ & $(148.4_{-1.3}^{+3.6})$ & $(147.8_{-3.8}^{+1.3})$ \\
\bottomrule
\end{tabular}}
\caption{Mean values and $68\%$ (95\%)~CL errors on the six $\Lambda$CDM parameters as well as on some derived ones ($H_0$, $\Omega_{\rm m}$, $\sigma_8$, $S_8$, $r_{\rm{drag}}$) after marginalizing over a complete and large number of possible fiducial cosmologies. We illustrate the results from Planck, ACT and SPT CMB observations, either alone or combined with low redshift measurements.}
\label{tab:marginalized_constraints}
\end{table*}

Tabs.~\ref{tab:marginalized_constraints} depicts the marginalized constraints (at both 68 and 95\% CL) on the different cosmological parameters explored here arising from the different data sets.
The bounds derived within each of the fiducial cosmologies described in the previous section are shown in Tabs.~\ref{ tab.results.lcdm }-\ref{ tab.results.w0wa }, which also include the minimal $\Lambda$CDM scenario.
In what follows, for each of the six main cosmological parameters, as well as other key derived quantities, we discuss the various correlations observed across different extended cosmological models, highlighting the main correlation and degeneracy patterns between standard and non-standard cosmological parameters. Ultimately, we present the marginalized posterior distribution functions, where information from all models converges, leading to constraints on the main cosmological parameters that should be considered more robust than those derived under a specific cosmological model. We also detail, case by case, the most important differences between the constraints obtained in our model-marginalized framework and those derived from individual models. Finally, we highlight the most relevant applications of our marginalized bounds on a parameter-by-parameter basis.

\begin{figure*}[htpb!]
\centering
\includegraphics[width=0.4\textwidth]{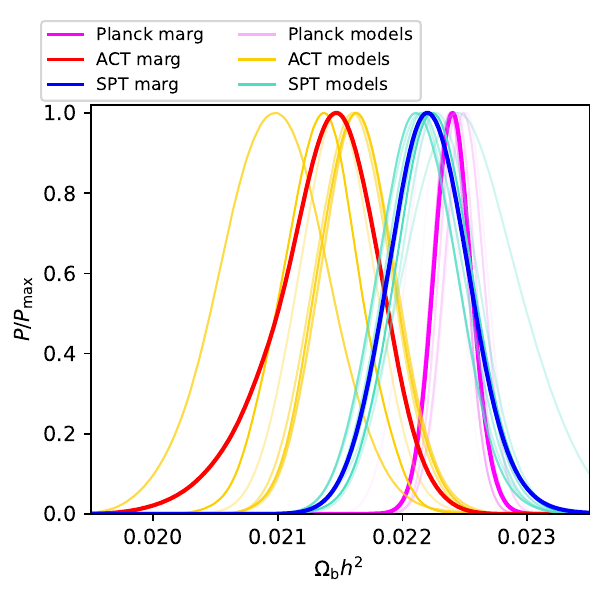}
\includegraphics[width=0.4\textwidth]{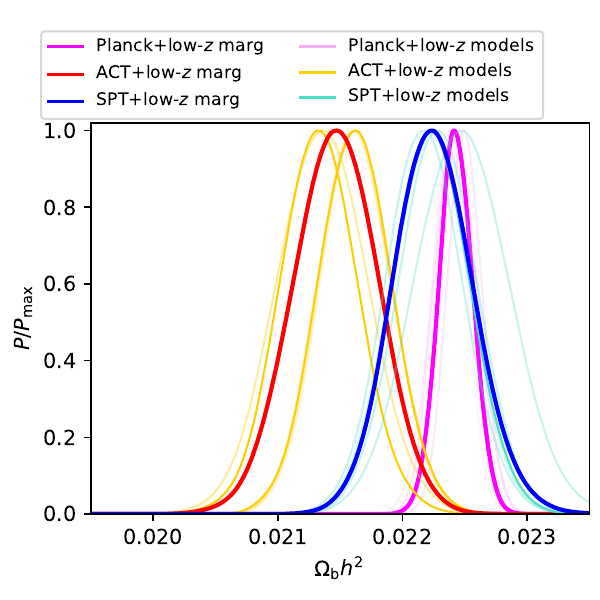}
\caption{\label{fig:marg_ombh2}
Marginalized 1D posteriors for $\Omega_\mathrm{b} h^2$, considering different datasets. Lighter colors (pink for Planck, yellow for ACT, light blue for SPT) indicate the posteriors for each of the various cosmological models, where the least favorite models correspond to fainter lines. The model-marginalized posteriors are shown in brighter colors (magenta for Planck, red for ACT, blue for SPT).
The left (right) panel represents results obtained without (with) low-$z$ data.
}
\end{figure*}

\subsection{Baryon Energy Density $\Omega_{\rm b} h^2$}

Let us start commenting on the baryon mass-energy density, $\Omega_{\rm b} h^2$. The subsection is organized as follows: we first discuss the main correlations between the baryon energy densities and the beyond-$\Lambda$CDM parameters. The aim of this subsection is twofold: \textit{(i)} on one hand, we want to understand what types of phenomenologies can alter the results on $\Omega_b h^2$ inferred within a minimal $\Lambda$CDM cosmology, thereby providing a robust physical interpretation of the reasons behind these changes; and \textit{(ii)} on the other hand, we want to identify the main sources of uncertainty surrounding the constraints on the baryon energy density obtained within the $\Lambda$CDM cosmology. Specifically, we will examine if and to what extent the constraining power on the baryon energy density depends on fixing other cosmological parameters beyond the standard six to their reference values expected within the baseline model of cosmology, and how relaxing these assumptions affects the constraints. Then, we adopt the Bayesian methodologies outlined in the previous sections to derive model-marginalized constraints on the baryon energy density for all the different combinations of datasets analyzed, incorporating information from all the minimal extended cosmologies under consideration. Finally, we comment on the overall consistency of the results and their applicability.

%Figure~\ref{fig:marg_ombh2} illustrates the shift in the one-dimensional posterior probability for $\Omega_{\rm b} h^2$ arising from all the fiducial cosmologies and all data combinations, including also the marginalized distributions.
%Notice firstly that the largest departure in the uncertainty of $\Omega_{\rm b} h^2$ from its reference value within the $\Lambda$CDM cosmology occurs when either $N_{\rm eff}$ or $A_{\rm lens}$ are free parameters, being twice as large for the effective number of neutrino species. 
\subsubsection{Main correlations with beyond $\Lambda$CDM parameters}
When comparing the results for $\Omega_b h^2$ obtained within the different fiducial extended cosmologies across the six combinations of data analyzed, we observe that the largest increase in the uncertainty of $\Omega_{\rm b} h^2$ from its reference value within the $\Lambda$CDM cosmology occurs when either $N_{\rm eff}$ or $A_{\rm lens}$ is treated as a free parameter, with the uncertainty being twice as large when varying the effective number of neutrino species. For these extended models, we show the 1-dimensional posterior distribution functions and the 2-dimensional marginalized contours (for all six combinations of data) in the top two panels of Fig.~\ref{fig:corr_main_6} (with the top left panel showing the joint contours in the $\Omega_b h^2$ - $A_{\rm lens}$ plane, and the top right panel showing the correlation in the $\Omega_b h^2$ - $\Delta N_{\rm eff}$ plane).

The correlation between $N_{\rm eff}$ and  $\Omega_b h^2$ was somehow expected, as the baryon mass energy density is measured from observations of the relative height of the second CMB peak with respect to the first and third peaks: if one adds baryons, the odd peaks are enhanced over the even peaks, that is, baryons make the first acoustic peak much larger than the second. The more baryons, the more the second peak is relatively suppressed, as the extra gravity provided by the baryons will enhance the compression into the potential wells. On the other hand, a larger $N_{\rm eff}$ will mostly affect the CMB power spectrum at high multipoles $\ell$, rather than at the very first peaks, i.e.\ at the CMB damping tail~\cite{Hou:2011ec}.  If $\Delta N_{\rm eff}$ increases, the Hubble parameter $H(z)$ during radiation domination will increase as well. Baryon-photon decoupling is not an instantaneous process, leading to a diffusion damping of oscillations in the plasma. If decoupling starts at $\tau_{\rm d}$ and ends at $\tau_{\rm ls}$, during $\Delta\tau$
the radiation free streams on scale $\lambda_{\rm d}=\left(\lambda\Delta\tau\right)^{1/2}$ where $\lambda$ is the photon mean free path and $\lambda_{\rm d}$ is shorter than the thickness of the last scattering surface. As a consequence, temperature fluctuations on scales smaller than $\lambda_{d}$ are damped, because on such scales photons can spread freely both from overdensities and underdensities. The overall result is that the damping angular scale $\theta_{\rm d}=r_{\rm d}/D_{\rm A}$ is proportional to the square root of the expansion rate $\theta_{\rm d}\propto\sqrt{H}$ and consequently it increases with $\Delta N_{\rm eff}$, inducing a suppression of the peaks located at high multipoles and a smearing of the oscillations that intensifies at the CMB damping tail. Therefore, when considering $N_{\rm eff}$ as a free parameter, the bounds on the baryon energy density are loosened as they affect the CMB anisotropies in nearby multipole regions. 

As seen in Fig.~\ref{fig:corr_main_6}, a similar situation, albeit to a minor extent, happens with the lensing amplitude $A_{\rm lens}$. CMB temperature fluctuations get blurred due to gravitational lensing by the large scale structure of the Universe: photons from different directions are mixed and the peaks at large multipoles are smoothed. The amount of lensing is a precise prediction of the $\Lambda$CDM model: the consistency of the model can be checked by artificially increasing lensing by a factor $A_{\rm{lens}}$~\cite{Calabrese:2008rt} (\emph{a priori} an unphysical parameter). If $\Lambda$CDM consistently describes all CMB data, observations should prefer $A_{\rm{lens}}=1$.  Since the effect of the lensing amplitude is also focused in the large multipoles peaks, it partly affects the extraction of $\Omega_{\rm b} h^2$. These larger uncertainties in $\Omega_{\rm b} h^2$ are observed in case of Planck, ACT and SPT data. Once low redshift measurements are accounted for, the uncertainties are reduced and are much closer to the standard $\Lambda$CDM ones. %Indeed, the marginalized errors on $\Omega_{\rm b} h^2$ depicted in Tab.~\ref{tab:marginalized_constraints_95} are very close to the $\Lambda$CDM ones, see Tab.~\ref{ tab.results.lcdm }. 

\subsubsection{Marginalized Constraints}
%Figure~\ref{fig:marg_ombh2} illustrates the shift in the one-dimensional posterior probability for $\Omega_{\rm b} h^2$ arising from all the fiducial cosmologies and all data combinations, including also the marginalized distributions. It confirms the effects described in the previous section: the spread in the mean values of $\Omega_{\rm b} h^2$ for ACT and SPT is reduced when low-z data are added in the numerical analyses. It also shows that the mean value preferred by ACT is lower than that preferred by SPT or Planck, showing a (mild) $\sim 2\sigma$ discrepancy. Interestingly, this mismatch does not dilute when adding low-redshift observations. The reason for that is due to the higher CMB second peak amplitude preferred by ACT measurements when compared to Planck observations, implying therefore a lower baryon mass energy density with respect to that preferred by Planck data. 
In Figure~\ref{fig:marg_ombh2}, we show the displacement in the one-dimensional posterior probability for $\Omega_{\rm b} h^2$ arising from assuming all the different fiducial cosmologies and all data combinations analyzed in this study. In the very same figure, we highlight in bold the marginalized distribution resulting from marginalizing across the different models. The results confirm the effects described in the previous subsection: specifically, the spread in the mean values of $\Omega_{\rm b} h^2$ for ACT and SPT is reduced when low-$z$ data are added in the numerical analyses. It also shows that the mean value preferred by ACT is lower than that preferred by SPT or Planck, indicating a (mild) $\sim 2\sigma$ discrepancy. Interestingly, this mismatch does not diminish when adding low-redshift observations. This is due to the higher CMB second peak amplitude preferred by ACT measurements compared to Planck observations, implying a lower baryon mass-energy density relative to that preferred by Planck data. We also note that the inclusion of low-redshift measurements does not lead to a significant reduction in the constraining power; for all the data combinations involving the three different CMB experiments, the constraints remain of the same order of magnitude. That said, we observe that the constraints based on Planck CMB temperature and polarization measurements remain a factor of two more constraining compared to those based on ACT and SPT. Overall, the marginalized constraints on $\Omega_{\rm b} h^2$ can be regarded as more robust compared to those derived within the baseline $\Lambda$CDM cosmology, as they account for marginalization over a wide range of cosmological models, many of which recast the presence of new physics prior to recombination, such as the value of $\Delta N_{\rm eff}$. Therefore, they can serve as a resilient reference value, for instance in studies aimed at testing late-time cosmologies. Typically, in these cases, a prior on $\Omega_{\rm b} h^2$ is adopted, either tracking the results obtained by Planck under the assumption of a $\Lambda$CDM cosmology or -- to reduce model dependence -- from the BBN prediction (which, however, increases the uncertainties compared to the CMB-based constraints). Since our results account for many $\Lambda$CDM extensions and possible effects, they can be considered more robust than those derived under the assumption of the minimal $\Lambda$CDM scenario and at least competitive in terms of model dependence with the prior resulting from BBN. Therefore, they can be used in those kinds of analyses.

\subsection{Cold Dark Matter Energy Density $\Omega_\mathrm{c} h^2$}

\begin{figure*}[htpb!]
\centering
\includegraphics[width=0.4\textwidth]{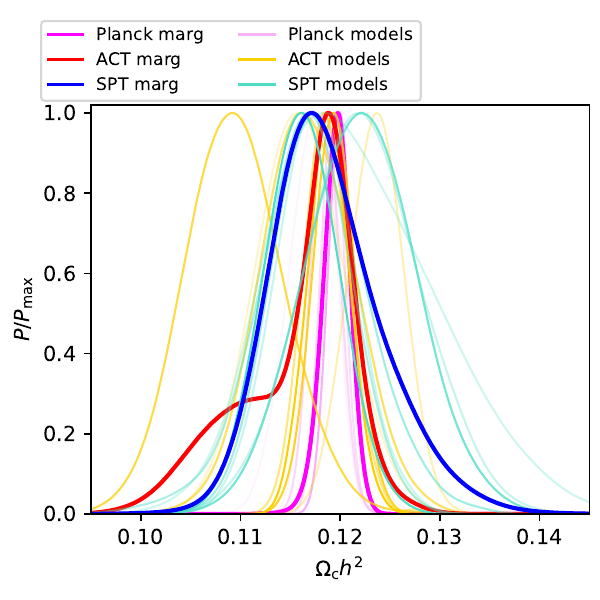}
\includegraphics[width=0.4\textwidth]{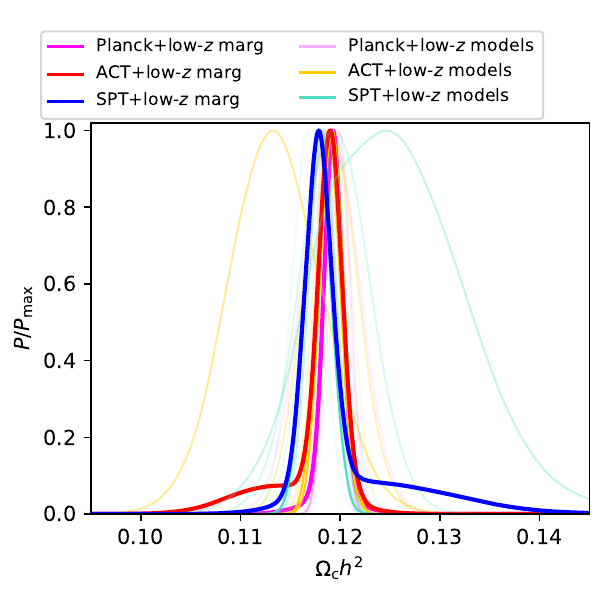}
\caption{\label{fig:marg_omch2}
Marginalized 1D posteriors for $\Omega_\mathrm{c} h^2$, considering different datasets.
Colors are the same as in Fig.~\ref{fig:marg_ombh2}.
}
\end{figure*}

The next parameter to be discussed is the dark matter mass-energy density. As usual, we start by scrutinizing the main correlations with non-standard cosmological parameters beyond the $\Lambda$CDM cosmology. We then derive marginalized constraints on this quantity, accounting for a wide range of minimal extensions across all six combinations of data.
 
\subsubsection{Main correlations with beyond $\Lambda$CDM parameters}
%As in the case of $\Omega_{\rm b}h^2$,  the largest departure from the canonical $\Lambda$CDM theory is also reached when either $N_{\rm eff}$ or $A_{\rm lens}$ are free parameters in the fiducial cosmology.
When it comes to correlations among parameters, the situation for the cold dark matter energy density turns out to be very similar to that of $\Omega_{\rm b} h^2$. In this case, the largest departure from the canonical $\Lambda$CDM theory is also observed when either $N_{\rm eff}$ or $A_{\rm lens}$ are free parameters in the fiducial cosmology. For all the combinations of data explored in this study, the one-dimensional posterior distribution functions and the two-dimensional correlations between $\Omega_{\rm c} h^2$, $N_{\rm eff}$, and $A_{\rm lens}$ are shown in the second row from the top in Fig.~\ref{fig:corr_main_6}. The left panel displays the correlation between $\Omega_{\rm c} h^2$ and $A_{\rm lens}$, while the right panel shows the correlation between $\Omega_{\rm c} h^2$ and $N_{\rm eff}$.

The degeneracy between $\Omega_{\rm c} h^2$ and $N_{\rm eff}$ can be easily understood in terms of the early Integrated Sachs Wolfe (ISW) effect. This effect is originated from the interaction between CMB photons and the time-dependent gravitational potentials along the line of sight between us and the last scattering surface. Notice that the early ISW effect adds in phase with the primary anisotropy, increasing the height of the first acoustic peaks, with an
emphasis on the first one,  where the main contribution of the ISW effect lies. In addition, the early ISW effect will be suppressed by the square of the radiation-to-matter ratio $\propto [(1 + z_{\rm r} )/(1 + z_{\rm eq} )]^2$, i.e.\ a larger (smaller) matter component will result into a smaller (larger) ISW amplitude due to the larger (smaller) value of $z_{\rm r}$. Conversely, a larger (smaller) amount of radiation -- i.e.\ larger (smaller) $N_{\rm eff}$ -- will result into a larger (smaller) ISW effect. This is also the reason for the lower value of $\Omega_\mathrm{c} h^2$ preferred by ACT observations, due to their slightly different amplitude of the first CMB acoustic peak when compared to Planck data.  

Concerning $A_{\rm lens}$, notice that there is always a preference for $A_{\rm lens}>1$ when considering Planck or SPT data, see Tab.~\ref{ tab.results.Alens }. This implies more CMB lensing and to compensate for that a lower matter density would be required. In the case of SPT, however, we find a value of $A_{\rm lens}<1$ and therefore the value of $\Omega_\mathrm{c} h^2$ will be higher than in the standard $\Lambda$CDM picture (see also Fig.~\ref{fig:corr_main_6}).  

\subsubsection{Marginalized Constraints}
%Figure~\ref{fig:marg_omch2} depicts the  marginalized 1D posteriors for $\Omega_\mathrm{c} h^2$, considering different datasets. Finally, the marginalized constraints on the dark matter energy density are very close to those obtained within the minimal scenario, reassuring the stability of the main cosmological parameters, despite the fact that also a time-dependent dark energy equation of state, parameterized via the CPL prescription, or a non-zero curvature, will also have a non-negligible impact on $\Omega_{\rm c}h^2$, see Tabs.~\ref{ tab.results.w0wa } and \ref{ tab.results.kcdm }. 
Figure~\ref{fig:marg_omch2} depicts the displacement in the one-dimensional posterior distribution functions for $\Omega_\mathrm{c} h^2$ resulting from assuming different cosmologies, along with the marginalized 1D posteriors highlighted in bold, considering various datasets. Overall, the marginalized constraints on the dark matter energy density are very close to those obtained within the minimal $\Lambda$CDM scenario, reassuring the stability of the main cosmological parameters. The constraints obtained from the different CMB datasets agree quite well with each other, both when including and excluding low-redshift observations.  However, we note that including low-redshift data is critical for achieving strong constraining power in the marginalized constraints. In contrast to the constraints on the baryon density discussed in the previous section (where the constraining power was relatively insensitive to the inclusion of low-redshift data), here incorporating local universe information can lead to an improvement in the resulting constraining power of more than a factor of two. This is due to the fact that a time-dependent dark energy equation of state, parameterized via the CPL prescription, or a non-zero curvature will also have a significant impact on $\Omega_{\rm c} h^2$ (see Tabs.~\ref{ tab.results.w0wa } and \ref{ tab.results.kcdm }). These parameters, which affect the late-time evolution of the universe, are almost unconstrained without distance observations but are well constrained when we add BAO and SN data.

\begin{figure*}[htpb!]
\centering
\includegraphics[width=0.4\textwidth]{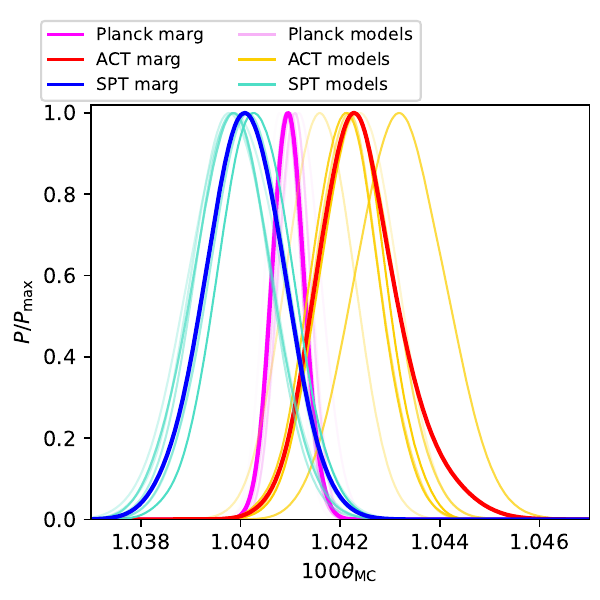}
\includegraphics[width=0.4\textwidth]{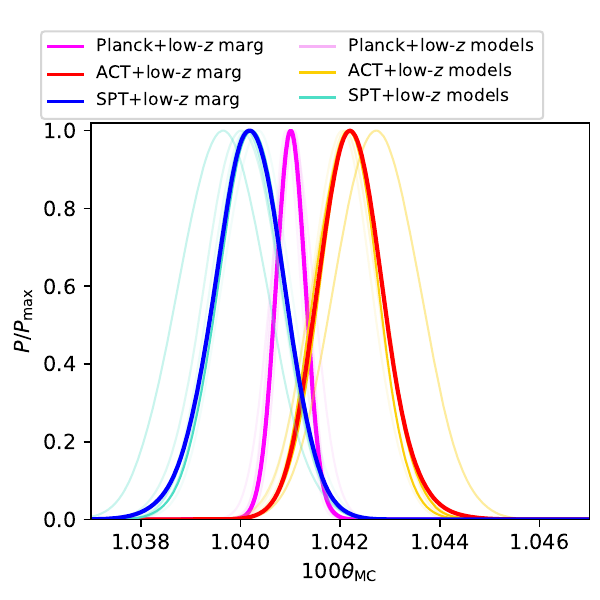}
\caption{\label{fig:marg_100theta}
Marginalized 1D posteriors for $100\theta_\mathrm{MC}$, considering different datasets.
Colors are the same as in Fig.~\ref{fig:marg_ombh2}.
}
\end{figure*}

\subsection{The Angular scale of the Sound Horizon, $\theta_{\rm MC}$}
The next parameter we shall discuss is $\theta_{\rm MC}$, an approximation to the angular scale of the sound horizon at decoupling, based on some model-dependent analytical fits.

\subsubsection{Main correlations with beyond $\Lambda$CDM parameters}

Looking at the constraints for the $\theta_{\rm MC}$ parameter in Tabs.~\ref{ tab.results.lcdm }–\ref{ tab.results.w0wa } for the different models, the first thing to notice is that $\theta_{\rm MC}$ is very well constrained by Planck CMB measurements. In contrast, higher multipole damping-tail CMB probes like SPT and ACT result in much larger errors. This is because Planck data has extremely high accuracy around the first acoustic peaks, while ACT and SPT focus on the high multipole region, where determining the angular position of the peaks is more challenging due to their smaller amplitudes. Additionally, low-redshift observations from BAO data do not significantly improve the constraints on $\theta_{\rm MC}$.

When varying the fiducial cosmology by allowing extensions beyond the minimal $\Lambda$CDM model, the impact on the mean value and errors of $\theta_{\rm MC}$ is generally small. The only notable exception is when $N_{\rm eff}$ is varied, as shown in Tab.~\ref{ tab.results.lcdm_nnu }. In Fig.~\ref{fig:corr_main_6} (third row, right panel), we plot the correlations between $N_{\rm eff}$ and $\theta_{\rm MC}$ for the different CMB experiments, both individually and in combination with low-redshift data. As seen in the figure, there is a quite strong negative correlation, especially in the ACT and SPT-based constraints. Increasing (or decreasing) the effective number of relativistic particles decreases (or increases) the angular scale of the sound horizon.

This correlation can be understood by referring to the analytical expressions for the sound horizon at decoupling, as discussed in Ref.~\cite{Kosowsky:2002zt}. Specifically, $r_s$, the sound horizon at decoupling, is directly linked to the amount of radiation in the universe, which is governed by $N_{\rm eff}$: increasing the radiation energy density decreases the size of the sound horizon. In turn, $100\theta_{\rm MC}$ is essentially the ratio of the sound horizon to the angular diameter distance to the CMB, which contains information about the universe’s expansion history after recombination. If we don’t change the late-time cosmology but only vary $N_{\rm eff}$ (or use low-redshift observations that constrain the post-recombination expansion to be very close to the Planck $\Lambda$CDM cosmology), this affects the numerator of this ratio without significantly changing the denominator. This has a clear impact on the inferred value of $100\theta_{\rm MC}$. This explains the relatively large effect that $N_{\rm eff}$ has on both the mean value and errors of $100\theta_{\rm MC}$, as shown in Fig.~\ref{fig:corr_main_6}.

\subsubsection{Marginalized Constraints}

As usual, in Fig.~\ref{fig:marg_100theta}, we depict the one-dimensional probability distributions for this parameter obtained from different combinations of datasets within the various fiducial cosmologies analyzed in this work. The bold lines represent the marginalized one-dimensional posterior probability distribution functions. The first noticeable point is that the displacement between the marginalized posterior and those obtained from different fiducial cosmologies is smaller than for other parameters, with the only exception being the results from varying $N_{\rm eff}$, as mentioned in the previous subsection.

The numerical results for $100\theta_{\rm MC}$ are summarized in Table~\ref{tab:marginalized_constraints}, both at the 68\% and 95\% CL. These results confirm that Planck-based constraints determine this parameter with a precision more than twice as accurate as those based on ACT and SPT. This is due to the lack of data from ACT and SPT around the first acoustic peak, which is crucial for precisely measuring the angular scale of the sound horizon.

Additionally, from Figs.~\ref{fig:marg_100theta}, we note that ACT-based constraints are shifted towards larger values of $\theta_{\rm MC}$ compared to Planck, while SPT-based constraints are shifted towards smaller values. This results in a mild (~$2\sigma$) difference in the constraints on this parameter from different CMB probes. The situation remains similar when considering constraints from the combination of CMB experiments and low-redshift data. However, in these cases, we observe a reduction in the displacement of the one-dimensional probability distributions (see the right panel of Fig.~\ref{fig:marg_100theta}) and a decrease in uncertainties for ACT and SPT-based marginalized constraints. On the other hand, the improvements for Planck-based constraints are minimal. This is because Planck data typically constrain this parameter well, even in extended cosmologies, due to the availability of data around the first peak. In contrast, ACT and SPT constraints allow more flexibility in shifting $\theta_{\rm MC}$ when studying fiducial cosmologies that modify the late-time expansion history. In these latter cases, the angular diameter distance to the CMB is well constrained only when low-redshift data are included, resulting in better-determined values for $\theta_{\rm MC}$.

We conclude this subsection with an important remark on the utility of our marginalized constraints. Specifically, the value of the angular scale of the sound horizon measured by Planck under the assumption of a fiducial $\Lambda$CDM cosmology is often used as a distance prior to compress CMB likelihoods and constrain extensions to the standard cosmological model. Essentially, the angular scale of the sound horizon can be viewed as an additional BAO data point at recombination (i.e., at $z\sim1090$) to be considered when testing new models. This parameter is commonly used because, as our analysis partially confirmed, $100\theta_{\rm MC}$ is considered a relatively model-independent geometric measurement characterizing the CMB angular power spectra (being closely tied to the position of the first acoustic peak). However, the marginalized result we provide in Table~\ref{tab:marginalized_constraints} already accounts for potential effects arising from varying a wide range of fiducial cosmologies considered here, thereby reducing the model dependence of our estimate. Therefore, this result can and should be used in such analyses instead of the most economical $\Lambda$CDM one, replacing the value obtained from assuming a single baseline cosmology only.
\begin{figure*}[htpb!]
\centering
\includegraphics[width=0.4\textwidth]{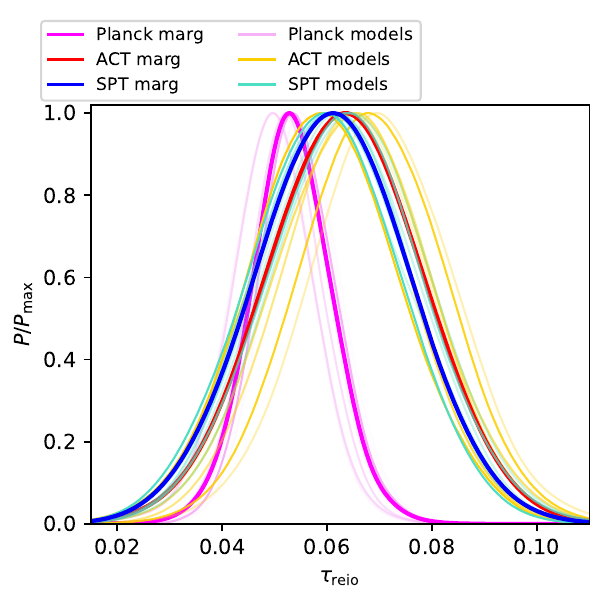}
\includegraphics[width=0.4\textwidth]{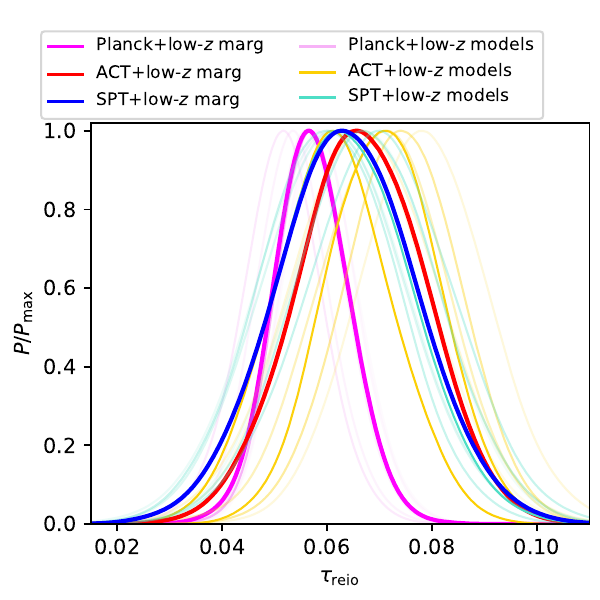}
\caption{\label{fig:marg_tau}
Marginalized 1D posteriors for $\tau_\mathrm{reio}$, considering different datasets.
Colors are the same as in Fig.~\ref{fig:marg_ombh2}.
In the ACT and SPT cases we are assuming a Gaussian prior, so these two experiments are not sensitive to $\tau_\mathrm{reio}$ and we are re-obtaining the prior used as an input. }
\end{figure*}

\subsection{Reionization optical depth, $\tau_{\rm {reio}}$}
We shall comment next on the correlations and marginalized limits of the parameter $\tau_{\rm {reio}}$, the reionization optical depth. 

\subsubsection{Main correlations with beyond $\Lambda$CDM parameters}

Constraints on $\tau_{\rm {reio}}$ based on Planck (and eventually ACT) CMB measurements, as well as its correlation with various beyond-$\Lambda$CDM parameters, have recently undergone extensive testing. Specifically, in Ref.~\cite{Giare:2023ejv}, some of us analyzed the reionization optical depth in relation to large-scale temperature and polarization measurements, investigating the possibility that anomalies in these parts of the datasets could lead to artificially low values of $\tau_{\rm {reio}}$, thereby driving several minor or mild anomalies in the Planck constraints. For more details in this direction, we refer to the aforementioned work. Here, we confirm that Planck’s accuracy on $\tau_{\rm {reio}}$ is largely driven by large-scale temperature and polarization measurements. In contrast, the errors from either the SPT or ACT datasets arise from the Gaussian prior applied when studying high-multipole CMB data from these two experiments, which, on their own, have no sensitivity to $\tau_{\rm {reio}}$. It’s worth noting that this Gaussian prior is a conservative estimate, partially derived from WMAP and Planck results (see Ref.~\cite{ACT:2020gnv}), and it results in errors twice as large as those obtained from Planck CMB measurements.\footnote{In Ref.~\cite{Giare:2023ejv}, some of us argued that, in principle, by combining low-redshift data with small-scale ACT CMB measurements, it is possible to constrain $\tau_{\rm {reio}}$ without adopting such a prior, thereby obtaining a Planck-independent estimate of this parameter. The resulting constraints tend to favor slightly larger values of the optical reionization depth, showing a $\sim 1.8\sigma$ tension with the Planck result. That being said, these constraints are significantly weaker in extended cosmological models, as demonstrated in the same work. This is the primary reason we chose to adopt the prior in our study.} Additionally, it should be noted that low-redshift measurements do not significantly alter the errors on $\tau_{\rm {reio}}$ obtained by Planck, due to their lack of sensitivity to the physical effects induced by this parameter.

Regarding the different fiducial cosmologies considered in this work, the most significant impact occurs, as expected, when the phenomenological parameter $A_{\rm lens}$ is allowed to vary freely (see Tab.~\ref{ tab.results.Alens } for numerical results and the third row, left panel of Fig.~\ref{fig:corr_main_6} for the one-dimensional posterior distributions and two-dimensional correlations between $\tau_{\rm {reio}}$ and $A_{\rm lens}$). As seen in the figure, due to this correlation, the mean value of $\tau_{\rm {reio}}$ is lower than its $\Lambda$CDM counterpart when the lensing amplitude is a free parameter in the analysis. Although the shift in the mean value is not significant, it is straightforward to understand: if $A_{\rm lens} > 1$, CMB lensing will be stronger. Since $\tau_{\rm {reio}}$ suppresses all acoustic peaks by a constant factor $\exp(-\tau_{\rm {reio}})$ while leaving the power on the largest scales unaffected, a smaller value of $\tau_{\rm {reio}}$ -- leading to reduced smearing of the small-scale features in the power spectrum -- will be required to compensate for this effect.

\begin{figure*}[htpb!]
\centering
\includegraphics[width=0.4\textwidth]{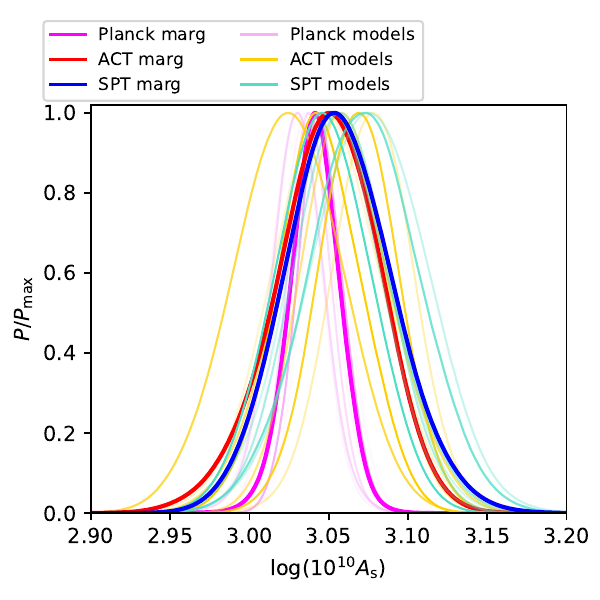}
\includegraphics[width=0.4\textwidth]{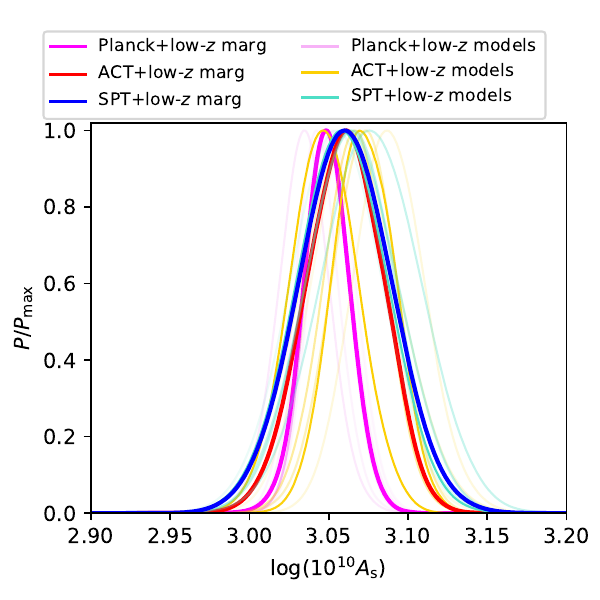}
\caption{\label{fig:marg_logA}
Marginalized 1D posteriors for $\log(10^{10} A_\mathrm{s})$, considering different datasets. Colors are the same as in Fig.~\ref{fig:marg_ombh2}.
}
\end{figure*}

\subsubsection{Marginalized Constraints}
The marginalized constraints on $\tau_{\rm {reio}}$ are shown in Table~\ref{tab:marginalized_constraints} and Fig.~\ref{fig:marg_tau}. In this case, we note that the constraints on the reionization optical depth are highly stable and robust. For the ACT and SPT-based results, this stability is primarily due to the inclusion of the Gaussian prior.  Minor shifts in the one-dimensional posterior distribution functions obtained within different extended cosmological models are observed, indicating that even with a prior on this parameter, slight displacements can arise when varying other quantities simultaneously. Our marginalized distributions (highlighted in bold in the figure) account for these minor shifts, weighting them by the Bayes factor of the different models.

\begin{figure*}[htpb!]
\centering
\includegraphics[width=0.4\textwidth]{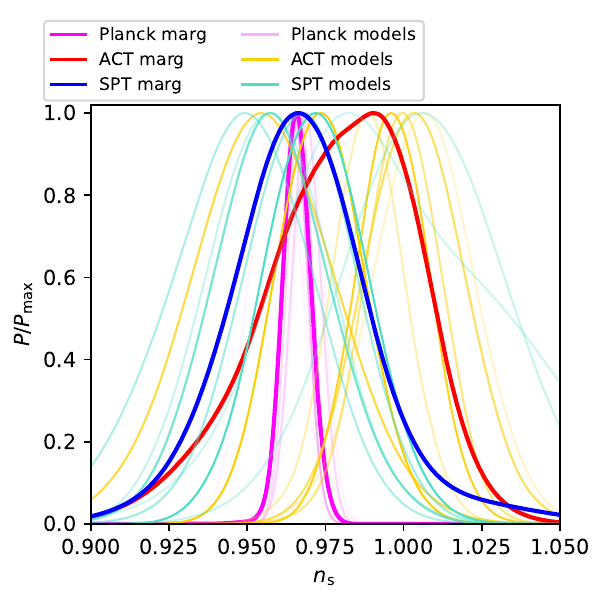}
\includegraphics[width=0.4\textwidth]{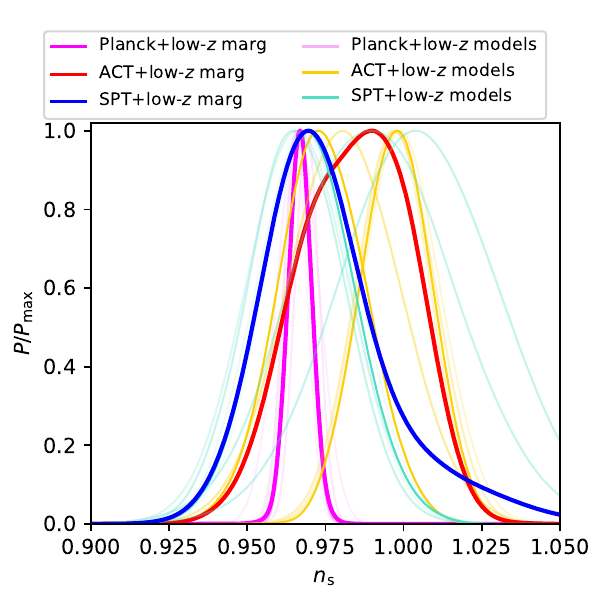}
\caption{\label{fig:marg_ns}
Marginalized 1D posteriors for $n_\mathrm{s}$, considering different datasets.Colors are the same as in Fig.~\ref{fig:marg_ombh2}.
}
\end{figure*}

\subsection{Amplitude of the primordial power spectrum $A_{\rm s}$}
Next, we shall focus on the amplitude of the primordial power spectrum $A_{\rm s}$.

\subsubsection{Main correlations with beyond $\Lambda$CDM parameters}
%Very similar results to those previously discussed for $\tau_{\rm{reio}}$ are obtained for the amplitude of the primordial power spectrum $A_{\rm s}$, strongly correlated with $\tau_{\rm reio}$: its most accurate measurements arise from Planck, low redshift data does not improve its determination and the largest impact on its error is when the lensing amplitude is a free parameter (see Tab.~\ref{ tab.results.Alens }), as $A_{\rm s}$ controls the overall normalization of the CMB power spectrum. 
When it comes to correlations between $\tau_{\rm{reio}}$ and beyond-$\Lambda$CDM parameters, we find very similar results to those discussed earlier for $\tau_{\rm{reio}}$. This is primarily due to the strong degeneracy between the amplitude of the primordial power spectrum, $A_{\rm s}$, and the optical reionization depth, $\tau_{\rm{reio}}$. A well-known effect of reionization, aside from enhancing the CMB polarization spectrum at large angular scales, is the suppression of temperature anisotropies at smaller scales. This suppression can be offset by increasing $A_{\rm s}$, which controls the overall normalization of the CMB power spectrum. Therefore, it is not surprising that the same parameters that correlate with $\tau_{\rm{reio}}$ also affect the amplitude of the primordial spectrum.

As shown in Tab.~\ref{ tab.results.Alens }, the largest impact on the determination of $A_{\rm s}$, particularly on its error, occurs when the lensing amplitude, $A_{\rm lens}$, is a free parameter. The one-dimensional posterior distribution functions and two-dimensional correlations for these parameters are presented in the bottom-left panel of Fig.~\ref{fig:corr_main_6}, and they exhibit a behavior similar to that described for the optical depth. Specifically, to counterbalance the effects of considering more (or less) lensing than expected in the standard $\Lambda$CDM model, one would need to decrease (or increase) the primordial amplitude. As seen in the figure and in Tab.~\ref{ tab.results.Alens }, low-redshift data do not significantly improve the constraining power in this case.

\subsubsection{Marginalized Constraints}
The marginalized constraints for $A_{\rm s}$ are summarized in Tab.~\ref{tab:marginalized_constraints} and visualized in the two panels of Fig.~\ref{fig:marg_logA}, where we also show the displacements in the one-dimensional probability distribution for this parameter across different datasets and fiducial cosmologies, together with the resulting marginalized one-dimensional posteriors in bold. Overall, the central values agree quite well across the various combinations of data, typically within one standard deviation. However, for CMB experiments other than Planck, the error budget is nearly doubled. This discrepancy in precision between Planck and Planck-independent experiments is somewhat mitigated by the inclusion of low-redshift data, although the uncertainty remains significantly larger than that for Planck combined with low-z data. In conclusion, low-redshift observations provide some improvement in reducing uncertainties in Planck-independent constraints, but Planck continues to offer superior precision in determining $A_{\rm s}$ across all cosmological models and datasets.

\begin{figure*}[htpb!]
\centering
\includegraphics[width=1\textwidth]{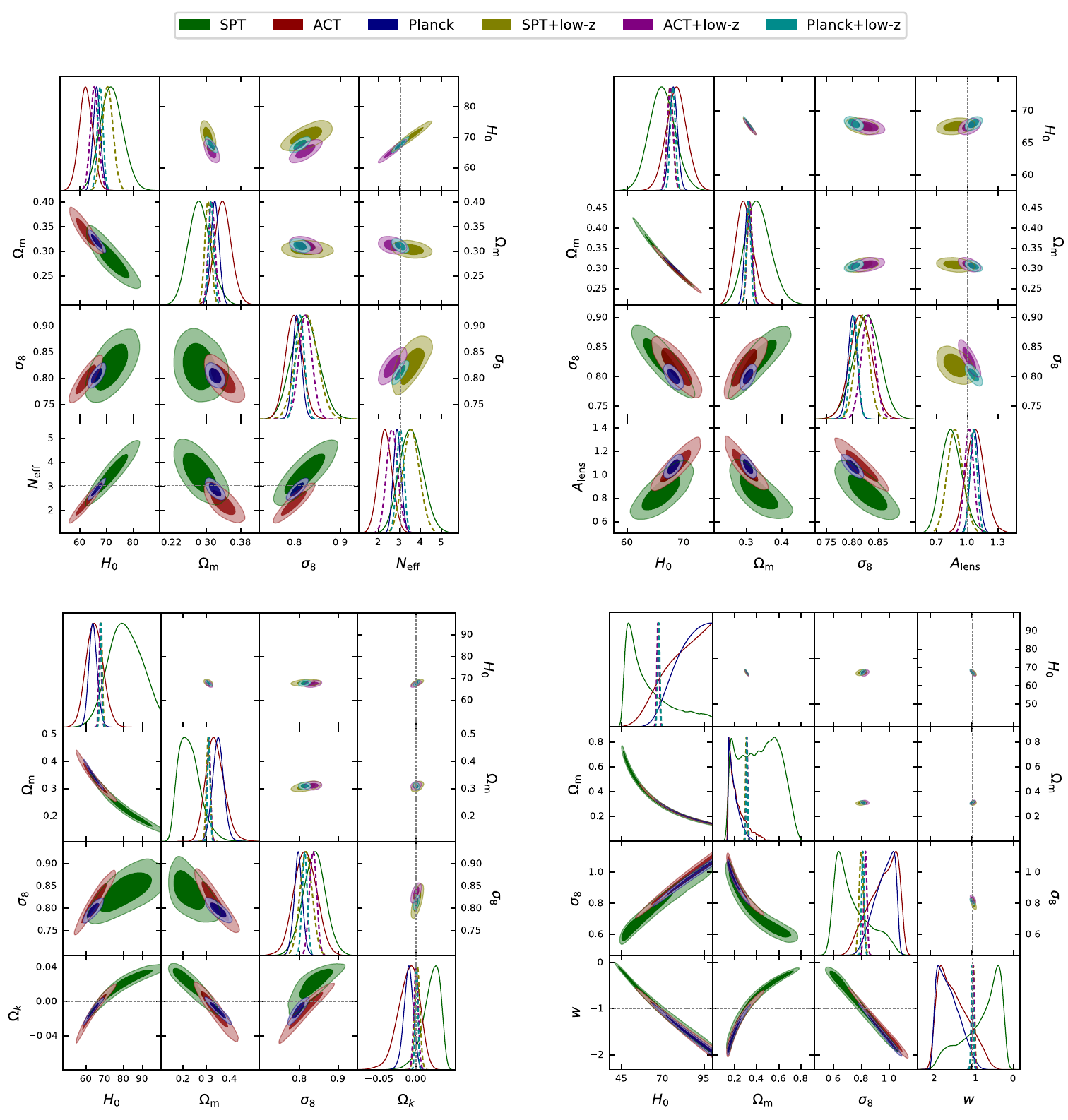}
\caption{Marginalized 1D and 2D posteriors involving the most relevant derived cosmological parameters ($H_0$, $\Omega_m$, $\sigma_8$) and other beyond-$\Lambda$CDM parameters that exhibit the most significant correlations and degeneracy lines with the former, across all the CMB and CMB+low-$z$ datasets analyzed in this work.}
\label{fig:corr_derived}
\end{figure*}

\begin{figure*}[htpb!]
\centering
\includegraphics[width=0.4\textwidth]{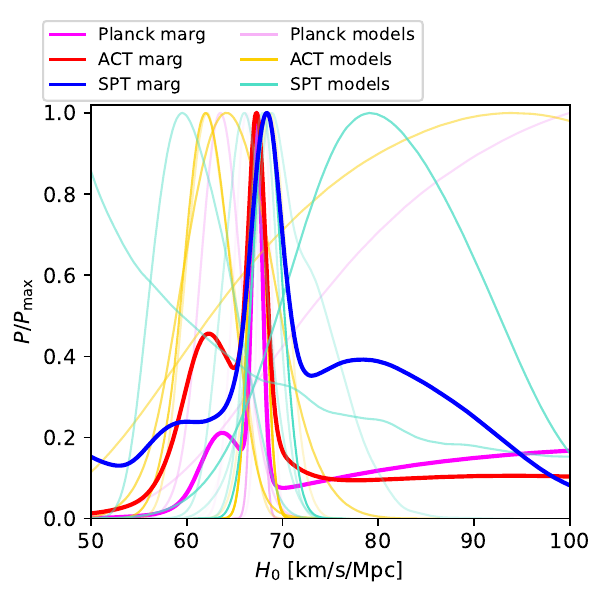}
\includegraphics[width=0.4\textwidth]{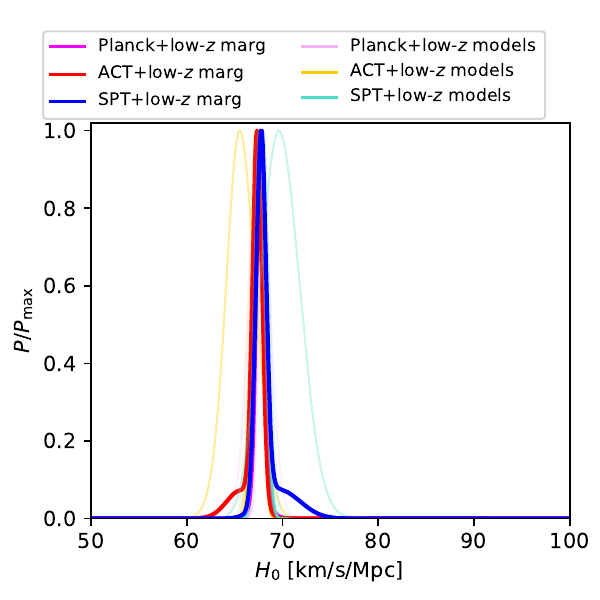}
\caption{\label{fig:marg_H0}
Marginalized 1D posteriors for $H_0$ [km/s/Mpc], considering different datasets.
Colors are the same as in Fig.~\ref{fig:marg_ombh2}.
}
\end{figure*}

\subsection{Primordial scalar spectral index $n_{\rm s}$}
The next parameter we discuss is the primordial power spectrum scalar spectral index, $n_{\rm s}$.  

\subsubsection{Main correlations with beyond $\Lambda$CDM parameters}

Constraints on the value of the spectral index of primordial fluctuations have recently been a matter of intense debate in the literature for several reasons. First and foremost, as highlighted by our extensive reanalysis, mild to moderate differences in the cosmological parameters inferred by ACT and Planck have emerged, as reported by several independent groups, including some of us, in Refs.~\cite{Lin:2019zdn,Forconi:2021que,Handley:2020hdp,LaPosta:2022llv,DiValentino:2022rdg,DiValentino:2022oon,Giare:2022rvg,Calderon:2023obf,Giare:2023xoc}. These discrepancies primarily involve the spectral index of primordial fluctuations, $n_s$, or inflationary parameters closely related to $n_s$ (see e.g., Refs.~\cite{Forconi:2021que, Giare:2022rvg, DiValentino:2022oon} for further discussion).  Unexpectedly, small-scale CMB measurements from the Data Release 4 of the Atacama Cosmology Telescope (ACT) show agreement with a scale-invariant Harrison-Zel'dovich spectrum ($n_s\sim 1$), introducing a tension with Planck results at 99.3\% CL ~\cite{Giare:2022rvg}. If observational systematics are set aside, and the differences between ACT and Planck are considered as genuine, the analysis of small-scale CMB data leads to distinct predictions for inflationary models~\cite{Giare:2023wzl}. Secondly, in Ref.~\cite{Giare:2024akf}, one of us extensively investigated how Planck's constraints on $n_s$ evolve in beyond-$\Lambda$CDM theoretical models aimed at resolving recent cosmological tensions (which we will discuss later in more detail). This analysis pointed out that deviations from a baseline $\Lambda$CDM cosmology at early times (i.e., before recombination) can significantly affect the results on the spectral index.

Unsurprisingly, our analysis confirms that the largest impact on the constraints on $n_s$ occurs when the number of relativistic degrees of freedom, $N_{\rm{eff}}$, is a free parameter in the fiducial cosmology (see Tab.~\ref{ tab.results.lcdm_nnu }). The correlations between $n_s$ and $N_{\rm{eff}}$ are shown in the bottom-right panel of Fig.~\ref{fig:corr_main_6}, where all datasets reveal a strong positive correlation between these two parameters. This effect is straightforward to understand: values of $N_{\rm{eff}}<3$ imply less damping in the high-multipole region of the CMB. A lower $N_{\rm{eff}}$ can be compensated by a lower $n_s$, which decreases the slope of the angular power spectrum, reducing the right side relative to the left one. Conversely, values of $N_{\rm{eff}}>3$ require the mean value of $n_s$ to increase to compensate for the larger damping induced by a larger dark radiation component. The figure also reveals that varying these two parameters simultaneously in the cosmological model tends to reduce the preference for $n_s \sim 1$ observed in ACT-based results. However, this comes at the cost of shifting $N_{\rm{eff}}$ towards slightly lower values than those expected within the standard model of cosmology and particle physics (see Ref.~\cite{DiValentino:2022rdg} for further discussion).

\subsubsection{Marginalized Constraints}
The marginalized constraints on the spectral index are summarized in Tab.~\ref{tab:marginalized_constraints} at 68\% and 95\% CL. As shown in Fig.~\ref{fig:marg_ns}, the scatter in the one-dimensional marginalized posteriors obtained from different fiducial cosmologies is quite significant. This underscores that the constraints on this parameter depend non-trivially on the overall cosmology, as previously discussed. Notably, the figure confirms an overall tendency for the posterior distributions based on ACT and SPT CMB measurements to prefer slightly larger values than those inferred from Planck CMB data, consistent with the references outlined earlier. In fact, the marginalized limit on $n_s$ from ACT data reported in Table~\ref{tab:marginalized_constraints} aligns with $n_s \sim 1$ and is significantly shifted compared to Planck results. SPT constraints are generally in line with Planck but have such large uncertainties that they remain consistent with ACT results as well. Among the three CMB experiments, the marginalized posterior distribution functions that most strongly rule out $n_s \sim 1$ are those based on Planck data. Even in this case, the error bars on the spectral index are significantly larger than those obtained within a baseline $\Lambda$CDM cosmology. Last but not least, the same figure and table illustrate that low-redshift measurements do not lead to substantial improvements when combined with Planck CMB data. Instead, they slightly enhance the accuracy of $n_s$ extraction when added to SPT or ACT observations.

\subsection{Hubble Constant, $H_0$}

We have discussed so far the main six $\Lambda$CDM parameters. Now we will focus on the derived ones, starting with the Hubble constant, $H_0$. 

\begin{figure*}[htpb!]
\centering
\includegraphics[width=0.4\textwidth]{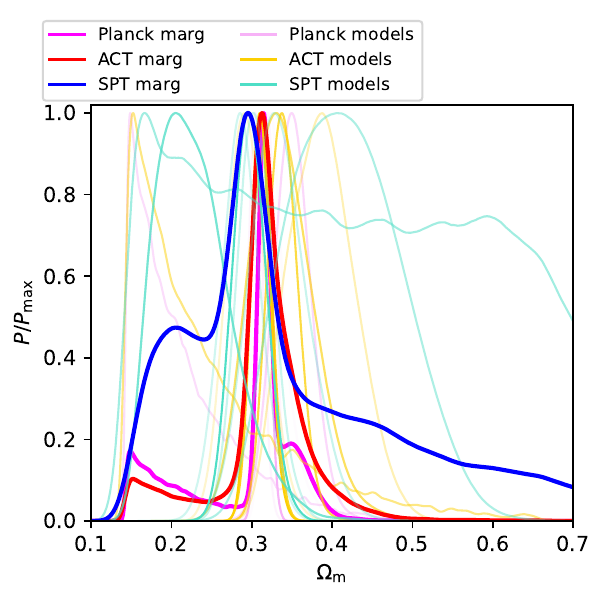}
\includegraphics[width=0.4\textwidth]{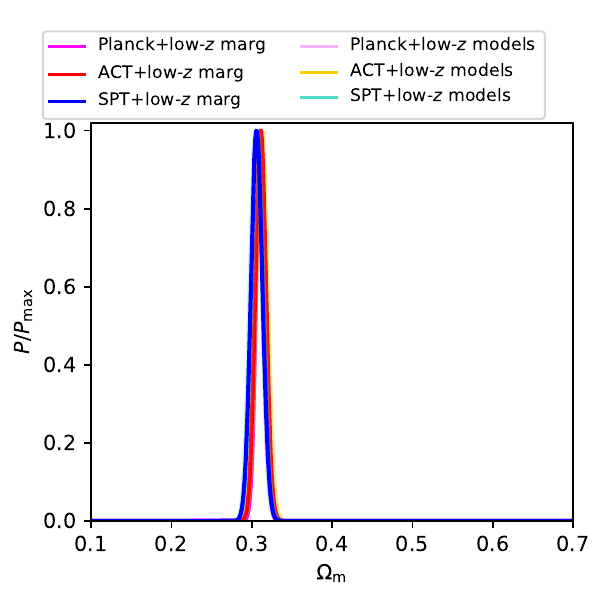}
\includegraphics[width=0.4\textwidth]{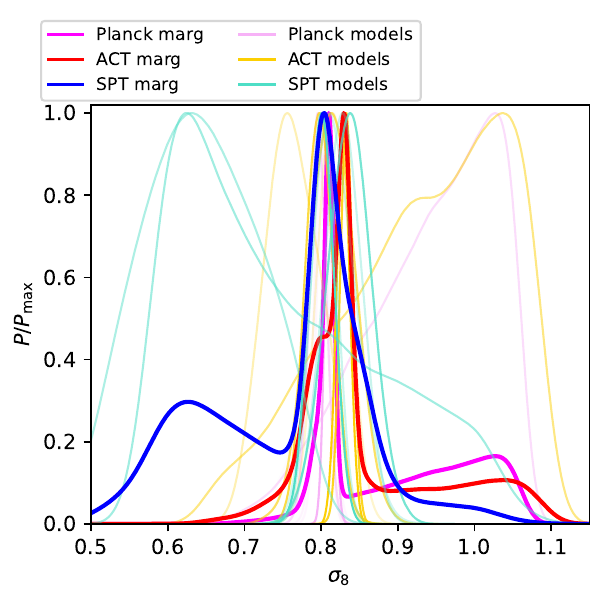}
\includegraphics[width=0.4\textwidth]{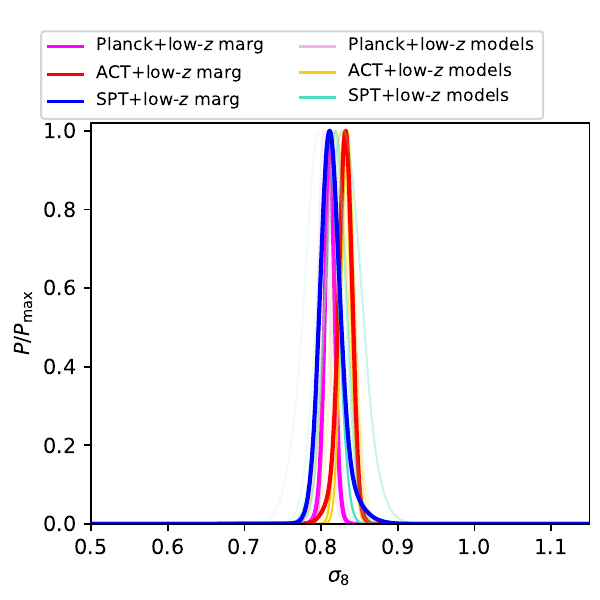}
\includegraphics[width=0.4\textwidth]{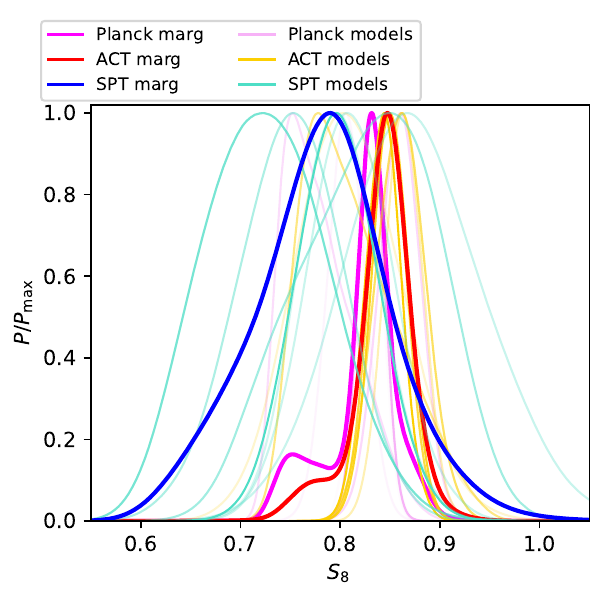}
\includegraphics[width=0.4\textwidth]{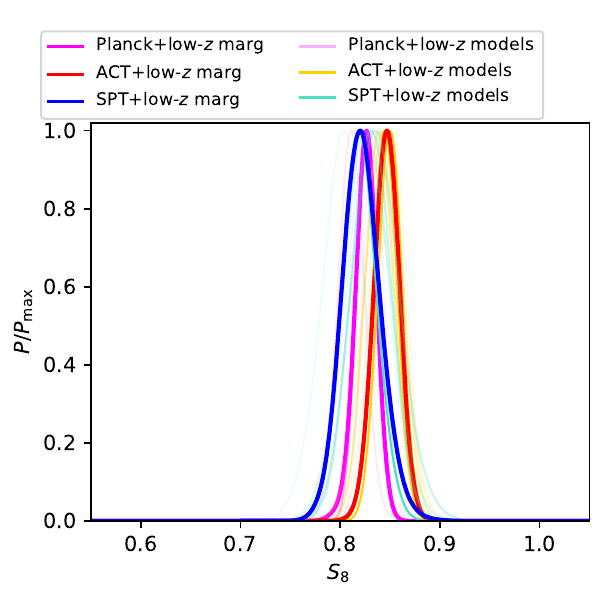}
\caption{\label{fig:marg_omegam}
Marginalized 1D posteriors for the matter energy density $\Omega_{\mathrm{m}}$ and the matter clustering parameters $\sigma_8$ and $S_8$, considering different datasets.Colors are the same as in Fig.~\ref{fig:marg_ombh2}.}
\end{figure*}

\subsubsection{Main correlations with beyond $\Lambda$CDM parameters}

When examining the correlation between $H_0$ and other cosmological parameters, the most significant increases in the errors and/or deviations from results inferred within a baseline $\Lambda$CDM cosmology occur when modifications are made to the dark energy sector.  As can be noticed from Tabs.~\ref{ tab.results.w } and \ref{ tab.results.w0wa }, a much larger value of the Hubble constant is obtained when the dark energy equation of state is a freely varying parameter or a function of redshift for the case of Planck and ACT. This is because of the well-known geometrical degeneracy among the equation of state, the matter density, and the Hubble constant.
As seen from the left bottom panel of Fig.~\ref{fig:corr_derived}, the value of the dark energy equation of state is preferred to be in the phantom region with a significance of $2\sigma$ (in the constant $w$ case). Therefore, changes in the properties or dynamics of dark energy can substantially impact the estimation of $H_0$ and its relationship with other parameters.
This is a well-known result and indeed it has been proposed as a scenario where to solve the long-standing Hubble constant tension~\cite{DiValentino:2016hlg}.  In the case of SPT, the situation is reversed: a value $w>-1$ is preferred (see the one dimensional probability distribution functions in Fig.~\ref{fig:corr_derived}), and therefore the value of the Hubble parameter is considerably smaller than within the minimal $\Lambda$CDM cosmology. Nevertheless, also in the SPT case the low redshift data restores mean values and errors of the Hubble constant very close to their $\Lambda$CDM counterparts. Overall, as largely expected, the inclusion of low-$z$ datasets becomes crucial when simultaneously varying parameters related to the dark sector and $H_0$. To emphasize this aspect, in Fig.~\ref{fig:corr_derived}, we have kept the same $y$-axis scale when plotting the two-dimensional marginalized contours with and without low-$z$ information to highlight the significant reduction in contour size compared to those obtained from CMB-only analyses.

Another significant shift of $H_0$ towards smaller values is instead due to the presence of massive neutrinos, that change the angular diameter distance at the last scattering surface and thus introduce a correlation with the Hubble constant. In this case, the shift for all the CMB experiments is in the same direction. Varying the neutrino effective number $N_{\rm eff}$ is again another method to relax the $H_0$ constraints, because of their very well known correlation through the effect on the peaks in the damping tail. The one-dimensional posterior distribution function and the two-dimensional correlation between $H_0$ and $N_{\rm eff}$ are shown in the top-left panel of Fig.~\ref{fig:corr_derived} for the different datasets. Also in this case, the final preferred value depends on the CMB observations under consideration: ACT points towards a lower value for $N_{\rm eff}$ and therefore a lower $H_0$ also, while SPT has the opposite trend. The very same argument can be used when the spatial curvature of the universe is free to vary. ACT goes in the direction of a closed universe and a lower $H_0$ value with error bars increasing up to a factor 5 with respect to the $\Lambda$CDM case, while SPT prefers an open universe and a higher $H_0$ value.

Last but not least, other important correlations involve the curvature parameter $\Omega_k$ and the lensing amplitude $A_{\rm lens}$, both of which are displayed in Fig.~\ref{fig:corr_derived} for the different datasets.

\subsubsection{Marginalized Constraints}

Table~\ref{tab:marginalized_constraints} shows the $95\%$~CL constraints on the Hubble parameter. While the mean value of $H_0$ is barely shifted, its errors are increased by one to two orders of magnitude when only CMB data are considered, depending on the dataset. This case is completely different from what we have observed so far, i.e., low redshift measurements are crucial for restoring the errors to their default values within the $\Lambda$CDM paradigm, see Tab.~\ref{ tab.results.lcdm }, Fig.~\ref{fig:marg_H0} and Fig.~\ref{fig:corr_derived}. Once low redshift probes are included in the analyses, the errors on $H_0$ are decreased by more than one order of magnitude. The very large spread in the errors of the Hubble constant is clearly visible from the left panel of Fig.~\ref{fig:marg_H0}. The right panel shows the dilution of such a spread due to BAO observations, which break many of the degeneracies between $H_0$ and the cosmological parameters involved in the extended cosmological scenarios considered here when only CMB observations are considered.

\begin{figure*}[htpb!]
\centering
\includegraphics[width=0.4\textwidth]{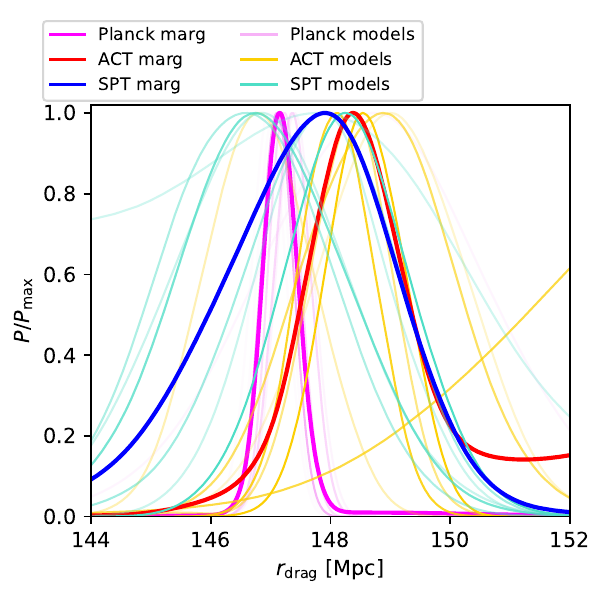}
\includegraphics[width=0.4\textwidth]{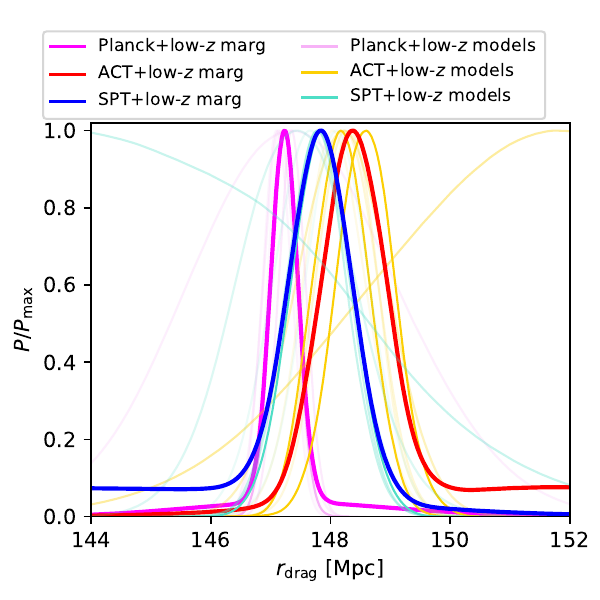}
\caption{\label{fig:marg_rdrag}
Marginalized 1D posteriors for $r_{\rm drag}$ [Mpc], considering different datasets.
Colors are the same as in Fig.~\ref{fig:marg_ombh2}.
}
\end{figure*}

\subsection{Matter Energy Density $\Omega_{\rm m}$ and Clustering parameters, $\sigma_8$ and $S_8$}

The following derived parameters we discuss are those related to the matter sector of the Universe, namely, the matter energy density, $\Omega_{\rm m}$, and the matter clustering-related parameters $\sigma_8$ and $S_8$.
%here

\subsubsection{Main correlations with beyond $\Lambda$CDM parameters}
The most relevant correlations involving the matter density and the clustering parameters are displayed in Fig.~\ref{fig:corr_derived}. In the top right panel, we show the correlation with the effective number of relativistic degrees of freedom. The top left panel highlights correlations with the lensing amplitude, while the bottom left explores the marginalized constraints with the curvature energy density. Finally, the bottom right panel shows the constraints with the dark energy equation of state. Among those, the largest shifts in both the matter density and matter clustering parameters occur when freedom is introduced into the dark energy sector. For datasets such as Planck and ACT, the Hubble constant can take higher values when $w<-1$. However, the structure of the CMB acoustic peaks restricts $\Omega_{\rm m} h^2$ from becoming too large. As a result, $\Omega_{\rm m}$ needs to be smaller than in the standard $\Lambda$CDM model. Conversely, in the case of SPT, a much larger value of $\Omega_{\rm m}$ is observed, which is almost unconstrained within the considered prior range. For a similar reason (i.e., its correlation with $H_0$), the matter density shifts to higher values when considering massive neutrinos, while the uncertainties relax by a factor of 2-3 compared to the $\Lambda$CDM scenario across all CMB datasets.
However, when low-redshift measurements are added for Planck, ACT, and SPT, the value of the dark energy equation of state approaches $w = -1$, bringing the mean values and uncertainties back in line with $\Lambda$CDM expectations. Low-redshift measurements also play a critical role in the constraints on $\sigma_8$ and $S_8$. Since low-redshift data directly measure these parameters, they help in breaking the degeneracies present in CMB-only analyses. Lastly, varying the neutrino effective number $N_{\rm eff}$ or the spatial curvature $\Omega_k$ results in significant shifts: ACT and Planck observations push the parameters towards higher values, while SPT tends to shift them towards lower values.

\subsubsection{Marginalized Constraints}
The marginalized constraints on the matter-related parameters are summarized in Tab.~\ref{tab:marginalized_constraints}. The posterior distribution functions for $\Omega_{\rm m}$, $\sigma_8$, and $S_8$, as well as the shifts in the posteriors derived from different models and datasets, are shown in Fig.~\ref{fig:marg_omegam}. In the left panel of Fig.~\ref{fig:marg_omegam}, we see that when considering only CMB data, there is significant scatter in the mean values and uncertainties of $\Omega_{\rm m}$, $\sigma_8$, and $S_8$. This is due to the challenge CMB data alone faces in constraining these quantities in extended cosmological models, largely because of correlations with other parameters, as discussed in the previous subsection. However, low-redshift measurements significantly reduce the uncertainties in the marginalized values. In some cases, they reduce the errors by an order of magnitude. The constraining power of low-$z$ data is evident in the right panel of the same figure, where the addition of these measurements narrows the scatter in the mean values and uncertainties of both the matter-energy density and the two clustering parameters, ensuring their stability against different possible fiducial cosmologies. As a result, the marginalized values of these derived parameters, especially when low-redshift data are included, offer a robust basis for testing exotic extensions of the $\Lambda$CDM scenario, due to their stability across various cosmological models.

\subsection{The sound horizon, $r_{\rm drag}$}
The last parameter we discuss is the value of the sound horizon at the drag epoch, $r_{\rm drag}$.\footnote{It is important to note that the sound horizon evaluated at the baryon drag epoch, referred to here as $r_{\rm drag}$, differs from the sound horizon evaluated at recombination because the two epochs are separated by $\Delta z \sim 30$. Since the results in the literature are commonly expressed in terms of $r_{\rm drag}$, we follow this convention.}

\subsubsection{Main correlations with beyond $\Lambda$CDM parameters}
By definition, the sound horizon contains information about the expansion history of the Universe from soon after inflation to before recombination. Therefore, when studying the correlation between the sound horizon and other cosmological parameters, it is no surprise that this parameter is primarily sensitive to quantities that influence the expansion history of the Universe before recombination. The main correlation we pointed out is with the effective number of relativistic degrees of freedom, $N_{\rm eff}$. Specifically, larger values of $N_{\rm eff} > 3$ will increase the radiation energy density of the Universe before recombination, leading to a faster expansion rate, $H(z)$. This, in turn, results in a smaller value for the sound horizon. Conversely, $N_{\rm eff} < 3$ has the opposite effect, leading to larger values of $r_{\rm drag}$. Due to its critical role in potentially addressing the long-standing Hubble tension, the correlation between the sound horizon and other cosmological parameters has been widely discussed in the literature. We have also highlighted these correlations ourselves in relation to other key cosmological parameters, such as $\theta_{\rm MC}$. Therefore, to avoid repetition of well-established information, we refer to previous subsections and the extensive literature dedicated to this topic for further details.

\subsubsection{Marginalized Constraints}
The marginalized constraints on the sound horizon at the drag epoch are summarized in Table~\ref{tab:marginalized_constraints} and visualized in Fig.~\ref{fig:marg_rdrag}. Several important observations are warranted here. Firstly, we note that the constraints achievable with Planck CMB observations are more than three times tighter compared to those based on ACT or SPT data. Including low-redshift information significantly improves the precision attainable with ACT and SPT, while it has a less pronounced impact on the Planck results. Consequently, when combining CMB data with local probes, the difference in constraining power between different datasets is reduced to a factor of 2. The improvement in the constraining power from local Universe probes is evident when comparing the left and right panels of Fig.~\ref{fig:marg_rdrag}. This can be explained by the sensitivity of the sound horizon to beyond-$\Lambda$CDM extensions. Despite this improvement, we observe that the marginalized uncertainties remain relatively large for experiments beyond Planck, and even within Planck they are broader than those inferred assuming a $\Lambda$CDM cosmology. 
This highlights how sensitive constraints on the sound horizon can be to changes in fiducial cosmology. This sensitivity includes not only parameters that affect early-time cosmology (such as $N_{\rm eff}$, discussed in the previous subsection) but also parameters influencing late-time expansion history, particularly for experiments other than Planck. This is evident from Fig.~\ref{fig:marg_rdrag}, where we observe significant scatter in the one-dimensional probability distribution functions obtained across different fiducial cosmologies.

We conclude with a final remark on the applicability of our marginalized constraints on the sound horizon. The value of $r_{\rm drag}$ inferred by Planck, assuming a baseline $\Lambda$CDM cosmology, is often adopted as a prior in studies that aim to compress the CMB likelihood when testing extended cosmological models (particularly when no changes are expected in the expansion history of the Universe before recombination). However, as highlighted by our analysis, mild deviations from this value are observed even when studying late-time expanded cosmologies due to correlations among parameters. Therefore, we argue that our marginalized constraints offer a safer and more robust choice for distance prior likelihoods. Our constraints are less model-dependent (although still based on extensions of the $\Lambda$CDM model) and more conservative, with larger uncertainties that better capture potential variations across different fiducial cosmologies. This makes them more suitable for extended model tests, avoiding the overly restrictive assumptions of a $\Lambda$CDM framework.

\section{Conclusions}
\label{sec.Conclusions}
In this work, we provide a comprehensive overview of the constraints on cosmological parameters derived from current CMB experiments, both individually and in combination with local Universe probes. We discuss the various correlations observed across different extended cosmological models on a parameter-by-parameter basis, highlighting key correlations and degeneracy patterns between standard and non-standard cosmological parameters and pointing out the most important differences among different datasets and CMB experiments. Ultimately, we present the marginalized posterior distribution functions, where information from all the considered extensions of the $\Lambda$CDM model converges, leading to constraints on the main cosmological parameters that are more robust than those derived from specific cosmological models.

Our study is motivated by several interesting reasons. First and foremost, we aim to address the following question: Are the $\Lambda$CDM model parameter values robust? The answer is yes. After analyzing the model-marginalized limits,
obtained considering several extensions of the simplest $\Lambda$CDM scenario, on the standard $\Lambda$CDM cosmological parameters $\Omega_{\rm c} h^2$, $\Omega_{\rm b} h^2$, $\theta_{\rm MC}$, $\tau_{\rm reio}$, $n_s$, and $A_s$, as well as on the derived quantities $H_0$, $\Omega_{\rm m}$, $\sigma_8$, $S_8$, and $r_{\rm drag}$, we find that, although the mean values and errors from CMB data alone have large spreads when computed over a number of possible non-minimal fiducial cosmologies, the addition of low-redshift measurements restores their values to the canonical $\Lambda$CDM ones.

Secondly, our marginalization procedure includes extensions of the minimal $\Lambda$CDM model with massive neutrinos, extra relativistic species, non-minimal dark energy equation of state scenarios, non-zero curvature, or a varying lensing amplitude parameter. Our results not only demonstrate the stability of the standard cosmological parameters within the minimal $\Lambda$CDM model and its extensions, but also provide a set of marginalized errors and mean values that are more reliable and conservative than those obtained within the minimal $\Lambda$CDM picture, serving as valuable tools for testing some extended cosmological scenarios.

In particular, we highlight the following findings and their potential applications. Our marginalized constraints on the baryon energy density parameter, $\Omega_{\rm b} h^2$, provide a more robust alternative to the traditional value inferred within $\Lambda$CDM cosmology, owing to their reasonably reduced model dependence. In studies focused on late-time cosmology, a prior on $\Omega_{\rm b} h^2$ is often adopted based on results from Planck under the $\Lambda$CDM assumption or, alternatively, from BBN predictions, which typically increase uncertainties compared to Planck-based constraints. Our results offer a reasonable alternative to both the BBN prior and the Planck $\Lambda$CDM-based result. Indeed, our marginalized constraint considers a broad range of models and possible effects and, although not entirely model-independent, can certainly be regarded as more conservative estimates than those derived under the $\Lambda$CDM assumption. They are also at least competitive with the BBN prior in terms of model dependence, making them effective for these analyses.

The Planck measurement of the angular scale of the acoustic peaks, $\theta_{\rm MC}$, is frequently used as a distance prior due to its relatively model-independent nature, being closely related to the position of the first acoustic peak in the CMB angular power spectra. $\theta_{\rm MC}$ can effectively serve as an additional BAO point at recombination (i.e., at $z \sim 1090$) when testing cosmological models and compressing CMB likelihoods. However, state-of-the-art analyses typically rely on the value inferred by Planck assuming a $\Lambda$CDM cosmology. Although this parameter is determined with a high level of precision and is not very sensitive to changes in cosmological models, minor deviations can nonetheless arise when considering various beyond-$\Lambda$CDM parameters. In contrast, our marginalized results account for potential effects arising from a variety of fiducial cosmologies, thereby further reducing the model dependence of our estimate. Thus, our marginalized constraints can and should be utilized in analyses, potentially replacing values obtained from single baseline cosmology assumptions, when a full analysis of cosmological data cannot be performed.

Finally, similarly to the previous parameters, the sound horizon constraint derived by Planck, based on the $\Lambda$CDM assumption, is often used as a prior in studies that aim to compress the CMB likelihood when testing extended cosmological models, particularly when no changes in the expansion history of the Universe are expected before recombination. However, our analysis reveals that mild deviations from this value can occur even when considering late-time expanded cosmologies due to parameter correlations. Consequently, we argue that our marginalized constraints represent a safer and more robust choice for distance prior likelihoods. These constraints are less model-dependent and more conservative, with larger uncertainties that better capture potential variations across cosmological models, thus avoiding the restrictive assumptions inherent in the $\Lambda$CDM framework.

\begin{acknowledgments}
\noindent
S.G.\
is supported by the European Union’s Framework Programme for Research
and Innovation Horizon 2020 (2014–2020) under Junior Leader Fellowship LCF/BQ/PI23/11970034 by La Caixa Foundation
and by
the Research grant TAsP (Theoretical Astroparticle Physics) funded by Istituto Nazionale di
Fisica Nucleare (INFN).
This work has been supported by the Spanish MCIN/AEI/10.13039/501100011033 grants PID2020-113644GB-I00 and by the European ITN project HIDDeN (H2020-MSCA-ITN-2019/860881-HIDDeN) and SE project ASYMMETRY (HORIZON-MSCA-2021-SE-01/101086085-ASYMMETRY) and well as by the Generalitat Valenciana grants PROMETEO/2019/083 and CIPROM/2022/69. OM acknowledges the financial support from the MCIU with funding from the European Union NextGenerationEU (PRTR-C17.I01) and Generalitat Valenciana (ASFAE/2022/020).
EDV is supported by a Royal Society Dorothy Hodgkin Research Fellowship. 
This article is based upon work from COST Action CA21136 Addressing observational tensions in cosmology with systematics and fundamental physics (CosmoVerse) supported by COST (European Cooperation in Science and Technology). 
We acknowledge IT Services at The University of Sheffield for the provision of services for High Performance Computing.
\end{acknowledgments}

%++++++++++++++++++++++++++
\newpage
\bibliography{biblio}	
%++++++++++++++++++++++++++
\newpage
\widetext
\appendix
\section{Tables}

\begin{table*}[h]
\begin{center}
\renewcommand{\arraystretch}{1.5}
\resizebox{\textwidth}{!}{
\begin{tabular}{l c c c c c c c c c c c c c c c }
\hline
\textbf{Parameter} & \textbf{ Planck } & \textbf{ Planck+low-z } & \textbf{ ACT } & \textbf{ ACT+lowz } & \textbf{ SPT } & \textbf{ SPT+low-z } \\ 
\hline\hline

$ \Omega_\mathrm{b} h^2  $ & $  0.02238\pm 0.00014\, ( 0.02238^{+0.00028}_{-0.00028} ) $ & $  0.02242\pm 0.00013\, ( 0.02242^{+0.00027}_{-0.00026} ) $ & $  0.02163\pm 0.00030\, ( 0.02163^{+0.00061}_{-0.00057} ) $ & $  0.02162\pm 0.00029\, ( 0.02162^{+0.00058}_{-0.00057} ) $ & $  0.02225\pm 0.00031\, ( 0.02225^{+0.00063}_{-0.00060} ) $ & $  0.02223\pm 0.00031\, ( 0.02223^{+0.00061}_{-0.00060} ) $ \\ 
$ \Omega_\mathrm{c} h^2  $ & $  0.1200\pm 0.0012\, ( 0.1200^{+0.0024}_{-0.0023} ) $ & $  0.11931\pm 0.00088\, ( 0.1193^{+0.0018}_{-0.0017} ) $ & $  0.1194\pm 0.0021\, ( 0.1194^{+0.0043}_{-0.0042} ) $ & $  0.1191\pm 0.0012\, ( 0.1191^{+0.0023}_{-0.0023} ) $ & $  0.1161\pm 0.0038\, ( 0.1161^{+0.0076}_{-0.0075} ) $ & $  0.1178\pm 0.0013\, ( 0.1178^{+0.0026}_{-0.0026} ) $ \\ 
$ 100\theta_\mathrm{MC}  $ & $  1.04091\pm 0.00031\, ( 1.04091^{+0.00061}_{-0.00061} ) $ & $  1.04100\pm 0.00029\, ( 1.04100^{+0.00056}_{-0.00057} ) $ & $  1.04209\pm 0.00068\, ( 1.0421^{+0.0013}_{-0.0014} ) $ & $  1.04213\pm 0.00061\, ( 1.0421^{+0.0012}_{-0.0012} ) $ & $  1.04029\pm 0.00075\, ( 1.0403^{+0.0015}_{-0.0015} ) $ & $  1.04022\pm 0.00068\, ( 1.0402^{+0.0013}_{-0.0013} ) $ \\ 
$ \tau_\mathrm{reio}  $ & $  0.0543\pm 0.0076\, ( 0.054^{+0.016}_{-0.015} ) $ & $  0.0576\pm 0.0071\, ( 0.058^{+0.015}_{-0.013} ) $ & $  0.069\pm 0.014\, ( 0.069^{+0.028}_{-0.027} ) $ & $  0.071\pm 0.011\, ( 0.071^{+0.021}_{-0.020} ) $ & $  0.059\pm 0.015\, ( 0.059^{+0.028}_{-0.029} ) $ & $  0.063\pm 0.013\, ( 0.063^{+0.026}_{-0.026} ) $ \\ 
$ n_\mathrm{s}  $ & $  0.9651\pm 0.0041\, ( 0.9651^{+0.0081}_{-0.0079} ) $ & $  0.9668\pm 0.0036\, ( 0.9668^{+0.0071}_{-0.0070} ) $ & $  0.996\pm 0.012\, ( 0.996^{+0.024}_{-0.024} ) $ & $  0.998\pm 0.012\, ( 0.998^{+0.022}_{-0.022} ) $ & $  0.972\pm 0.017\, ( 0.972^{+0.033}_{-0.033} ) $ & $  0.969\pm 0.015\, ( 0.969^{+0.030}_{-0.029} ) $ \\ 
$ \log(10^{10} A_\mathrm{s})  $ & $  3.044\pm 0.015\, ( 3.044^{+0.029}_{-0.028} ) $ & $  3.050\pm 0.014\, ( 3.050^{+0.028}_{-0.027} ) $ & $  3.069\pm 0.025\, ( 3.069^{+0.050}_{-0.050} ) $ & $  3.072\pm 0.019\, ( 3.072^{+0.037}_{-0.038} ) $ & $  3.046\pm 0.030\, ( 3.046^{+0.058}_{-0.060} ) $ & $  3.058\pm 0.029\, ( 3.058^{+0.055}_{-0.057} ) $ \\ 
$ H_0  $ & $  67.36\pm 0.54\, ( 67.4^{+1.1}_{-1.0} ) $ & $  67.65\pm 0.40\, ( 67.65^{+0.78}_{-0.78} ) $ & $  67.35\pm 0.90\, ( 67.4^{+1.8}_{-1.8} ) $ & $  67.43\pm 0.48\, ( 67.43^{+0.98}_{-0.92} ) $ & $  68.4\pm 1.5\, ( 68.4^{+3.1}_{-3.0} ) $ & $  67.78\pm 0.52\, ( 67.8^{+1.0}_{-1.0} ) $ \\ 
$ \Omega_\mathrm{m}  $ & $  0.3152\pm 0.0074\, ( 0.315^{+0.015}_{-0.014} ) $ & $  0.3111\pm 0.0053\, ( 0.311^{+0.011}_{-0.010} ) $ & $  0.313\pm 0.013\, ( 0.313^{+0.026}_{-0.024} ) $ & $  0.3110\pm 0.0065\, ( 0.311^{+0.013}_{-0.013} ) $ & $  0.298\pm 0.021\, ( 0.298^{+0.044}_{-0.040} ) $ & $  0.3062\pm 0.0071\, ( 0.306^{+0.014}_{-0.014} ) $ \\ 
$ \sigma_8  $ & $  0.8111\pm 0.0061\, ( 0.811^{+0.012}_{-0.012} ) $ & $  0.8116\pm 0.0059\, ( 0.812^{+0.012}_{-0.011} ) $ & $  0.8331\pm 0.0083\, ( 0.833^{+0.016}_{-0.016} ) $ & $  0.8343\pm 0.0078\, ( 0.834^{+0.015}_{-0.015} ) $ & $  0.800\pm 0.016\, ( 0.800^{+0.031}_{-0.032} ) $ & $  0.811\pm 0.012\, ( 0.811^{+0.023}_{-0.023} ) $ \\ 
$ r_\mathrm{drag}  $ & $  147.11\pm 0.27\, ( 147.11^{+0.53}_{-0.54} ) $ & $  147.22\pm 0.22\, ( 147.22^{+0.44}_{-0.44} ) $ & $  148.11\pm 0.63\, ( 148.1^{+1.2}_{-1.2} ) $ & $  148.18\pm 0.47\, ( 148.18^{+0.93}_{-0.90} ) $ & $  148.3\pm 1.1\, ( 148.3^{+2.1}_{-2.1} ) $ & $  147.86\pm 0.51\, ( 147.9^{+1.0}_{-1.0} ) $ \\ 

\hline \hline
\end{tabular} }
\end{center}
\caption{Mean values and $68\%$ ($95\%$)~CL constraints on the six $\Lambda$CDM parameters as well as on some derived ones ($H_0$, $\Omega_{\rm m}$, $\sigma_8$ and $r_{\rm{drag}}$) within the minimal $\Lambda$CDM fiducial cosmology.}
\label{ tab.results.lcdm }
\end{table*}

\begin{table*}[h]
\begin{center}
\renewcommand{\arraystretch}{1.5}
\resizebox{\textwidth}{!}{
\begin{tabular}{l c c c c c c c c c c c c c c c }
\hline
\textbf{Parameter} & \textbf{ Planck } & \textbf{ Planck+low-z } & \textbf{ ACT } & \textbf{ ACT+lowz } & \textbf{ SPT } & \textbf{ SPT+low-z } \\ 
\hline\hline

$ \sum m_\nu \, [eV]  $ & $ < 0.406 $ & $ < 0.135 $ & $ < 0.890 $ & $ < 0.207 $ & $ < 2.75 $ & $ < 0.242 $ \\ 
$ \Omega_\mathrm{b} h^2  $ & $  0.02231\pm 0.00016\, ( 0.02231^{+0.00031}_{-0.00031} ) $ & $  0.02243\pm 0.00013\, ( 0.02243^{+0.00026}_{-0.00026} ) $ & $  0.02147\pm 0.00031\, ( 0.02147^{+0.00059}_{-0.00060} ) $ & $  0.02163\pm 0.00029\, ( 0.02163^{+0.00058}_{-0.00056} ) $ & $  0.02218\pm 0.00032\, ( 0.02218^{+0.00062}_{-0.00061} ) $ & $  0.02224\pm 0.00031\, ( 0.02224^{+0.00061}_{-0.00061} ) $ \\ 
$ \Omega_\mathrm{c} h^2  $ & $  0.1206\pm 0.0013\, ( 0.1206^{+0.0026}_{-0.0025} ) $ & $  0.11918\pm 0.00089\, ( 0.1192^{+0.0017}_{-0.0017} ) $ & $  0.1236\pm 0.0028\, ( 0.1236^{+0.0054}_{-0.0054} ) $ & $  0.1185\pm 0.0012\, ( 0.1185^{+0.0024}_{-0.0025} ) $ & $  0.1161^{+0.0056}_{-0.0050}\, ( 0.116^{+0.011}_{-0.011} ) $ & $  0.1166\pm 0.0016\, ( 0.1166^{+0.0030}_{-0.0033} ) $ \\ 
$ 100\theta_\mathrm{MC}  $ & $  1.04080\pm 0.00032\, ( 1.04080^{+0.00063}_{-0.00061} ) $ & $  1.04101\pm 0.00029\, ( 1.04101^{+0.00055}_{-0.00056} ) $ & $  1.04159\pm 0.00068\, ( 1.0416^{+0.0013}_{-0.0013} ) $ & $  1.04222\pm 0.00062\, ( 1.0422^{+0.0012}_{-0.0012} ) $ & $  1.03984\pm 0.00075\, ( 1.0398^{+0.0015}_{-0.0015} ) $ & $  1.04032\pm 0.00069\, ( 1.0403^{+0.0013}_{-0.0013} ) $ \\ 
$ \tau_\mathrm{reio}  $ & $  0.0553\pm 0.0075\, ( 0.055^{+0.016}_{-0.014} ) $ & $  0.0589\pm 0.0074\, ( 0.059^{+0.015}_{-0.014} ) $ & $  0.070\pm 0.014\, ( 0.070^{+0.028}_{-0.028} ) $ & $  0.078\pm 0.012\, ( 0.078^{+0.025}_{-0.023} ) $ & $  0.061\pm 0.014\, ( 0.061^{+0.029}_{-0.028} ) $ & $  0.067\pm 0.014\, ( 0.067^{+0.026}_{-0.026} ) $ \\ 
$ n_\mathrm{s}  $ & $  0.9630\pm 0.0046\, ( 0.9630^{+0.0087}_{-0.0094} ) $ & $  0.9671\pm 0.0036\, ( 0.9671^{+0.0070}_{-0.0071} ) $ & $  0.989\pm 0.012\, ( 0.989^{+0.024}_{-0.024} ) $ & $  0.997\pm 0.012\, ( 0.997^{+0.023}_{-0.023} ) $ & $  0.947\pm 0.023\, ( 0.947^{+0.048}_{-0.050} ) $ & $  0.971\pm 0.015\, ( 0.971^{+0.030}_{-0.029} ) $ \\ 
$ \log(10^{10} A_\mathrm{s})  $ & $  3.048\pm 0.015\, ( 3.048^{+0.030}_{-0.028} ) $ & $  3.053\pm 0.015\, ( 3.053^{+0.029}_{-0.028} ) $ & $  3.078\pm 0.026\, ( 3.078^{+0.050}_{-0.050} ) $ & $  3.088\pm 0.023\, ( 3.088^{+0.046}_{-0.043} ) $ & $  3.058\pm 0.030\, ( 3.058^{+0.059}_{-0.058} ) $ & $  3.065\pm 0.029\, ( 3.065^{+0.057}_{-0.056} ) $ \\ 
$ H_0  $ & $  66.1^{+1.4}_{-1.2}\, ( 66.1^{+2.5}_{-2.8} ) $ & $  67.47\pm 0.43\, ( 67.47^{+0.83}_{-0.87} ) $ & $  62.4\pm 2.3\, ( 62.4^{+4.6}_{-4.4} ) $ & $  67.16\pm 0.53\, ( 67.2^{+1.0}_{-1.1} ) $ & $  60.8\pm 3.6\, ( 61^{+8}_{-7} ) $ & $  67.59\pm 0.54\, ( 67.6^{+1.1}_{-1.1} ) $ \\ 
$ \Omega_\mathrm{m}  $ & $  0.332^{+0.016}_{-0.020}\, ( 0.332^{+0.040}_{-0.035} ) $ & $  0.3132\pm 0.0057\, ( 0.313^{+0.011}_{-0.011} ) $ & $  0.388\pm 0.041\, ( 0.388^{+0.087}_{-0.085} ) $ & $  0.3136\pm 0.0069\, ( 0.314^{+0.014}_{-0.013} ) $ & $  0.416\pm 0.068\, ( 0.42^{+0.15}_{-0.14} ) $ & $  0.3071\pm 0.0072\, ( 0.307^{+0.014}_{-0.014} ) $ \\ 
$ \sigma_8  $ & $  0.788^{+0.026}_{-0.020}\, ( 0.788^{+0.046}_{-0.052} ) $ & $  0.8058\pm 0.0079\, ( 0.806^{+0.017}_{-0.017} ) $ & $  0.762\pm 0.034\, ( 0.762^{+0.065}_{-0.065} ) $ & $  0.824\pm 0.011\, ( 0.824^{+0.024}_{-0.025} ) $ & $  0.644\pm 0.076\, ( 0.64^{+0.14}_{-0.14} ) $ & $  0.793^{+0.021}_{-0.018}\, ( 0.793^{+0.038}_{-0.040} ) $ \\ 
$ r_\mathrm{drag}  $ & $  146.96\pm 0.30\, ( 146.96^{+0.57}_{-0.61} ) $ & $  147.24\pm 0.22\, ( 147.24^{+0.44}_{-0.43} ) $ & $  146.79\pm 0.90\, ( 146.8^{+1.7}_{-1.7} ) $ & $  148.32\pm 0.47\, ( 148.32^{+0.93}_{-0.93} ) $ & $  146.6\pm 1.4\, ( 146.6^{+2.7}_{-2.7} ) $ & $  148.12\pm 0.56\, ( 148.1^{+1.1}_{-1.1} ) $ \\ 

\hline \hline
\end{tabular} }
\end{center}
\caption{Mean values and $68\%$ ($95\%$)~CL constraints on the six $\Lambda$CDM parameters as well as on some derived ones ($H_0$, $\Omega_{\rm m}$, $\sigma_8$ and $r_{\rm{drag}}$) within a fiducial cosmology with massive neutrinos. The $95\%$~CL limits on the total neutrino mass are also presented, for the sake of completeness.}
\label{ tab.results.lcdm_mnu }
\end{table*}

\begin{table*}[h]
\begin{center}
\renewcommand{\arraystretch}{1.5}
\resizebox{\textwidth}{!}{
\begin{tabular}{l c c c c c c c c c c c c c c c }
\hline
\textbf{Parameter} & \textbf{ Planck } & \textbf{ Planck+low-z } & \textbf{ ACT } & \textbf{ ACT+lowz } & \textbf{ SPT } & \textbf{ SPT+low-z } \\ 
\hline\hline

$ N_\mathrm{eff}  $ & $  2.89\pm 0.19\, ( 2.89^{+0.38}_{-0.36} ) $ & $  3.03\pm 0.17\, ( 3.03^{+0.33}_{-0.33} ) $ & $  2.31\pm 0.34\, ( 2.31^{+0.68}_{-0.65} ) $ & $  2.72\pm 0.27\, ( 2.72^{+0.53}_{-0.51} ) $ & $  3.45^{+0.54}_{-0.63}\, ( 3.5^{+1.2}_{-1.1} ) $ & $  3.46\pm 0.38\, ( 3.46^{+0.84}_{-0.80} ) $ \\ 
$ \Omega_\mathrm{b} h^2  $ & $  0.02224\pm 0.00022\, ( 0.02224^{+0.00043}_{-0.00042} ) $ & $  0.02242\pm 0.00017\, ( 0.02242^{+0.00034}_{-0.00034} ) $ & $  0.02096\pm 0.00044\, ( 0.02096^{+0.00084}_{-0.00087} ) $ & $  0.02135\pm 0.00036\, ( 0.02135^{+0.00070}_{-0.00071} ) $ & $  0.02249\pm 0.00046\, ( 0.02249^{+0.00099}_{-0.00096} ) $ & $  0.02248\pm 0.00038\, ( 0.02248^{+0.00075}_{-0.00073} ) $ \\ 
$ \Omega_\mathrm{c} h^2  $ & $  0.1178\pm 0.0029\, ( 0.1178^{+0.0058}_{-0.0055} ) $ & $  0.1191\pm 0.0029\, ( 0.1191^{+0.0057}_{-0.0056} ) $ & $  0.1094\pm 0.0050\, ( 0.109^{+0.010}_{-0.0095} ) $ & $  0.1135\pm 0.0047\, ( 0.1135^{+0.0093}_{-0.0091} ) $ & $  0.1216^{+0.0084}_{-0.0094}\, ( 0.122^{+0.018}_{-0.017} ) $ & $  0.1253\pm 0.0072\, ( 0.125^{+0.016}_{-0.015} ) $ \\ 
$ 100\theta_\mathrm{MC}  $ & $  1.04117\pm 0.00044\, ( 1.04117^{+0.00086}_{-0.00085} ) $ & $  1.04104\pm 0.00043\, ( 1.04104^{+0.00085}_{-0.00083} ) $ & $  1.04321\pm 0.00089\, ( 1.0432^{+0.0018}_{-0.0017} ) $ & $  1.04275\pm 0.00083\, ( 1.0427^{+0.0016}_{-0.0016} ) $ & $  1.03992\pm 0.00092\, ( 1.0399^{+0.0018}_{-0.0018} ) $ & $  1.03963\pm 0.00086\, ( 1.0396^{+0.0017}_{-0.0017} ) $ \\ 
$ \tau_\mathrm{reio}  $ & $  0.0533\pm 0.0074\, ( 0.053^{+0.015}_{-0.014} ) $ & $  0.0574\pm 0.0072\, ( 0.057^{+0.015}_{-0.014} ) $ & $  0.060\pm 0.015\, ( 0.060^{+0.029}_{-0.029} ) $ & $  0.074\pm 0.011\, ( 0.074^{+0.021}_{-0.022} ) $ & $  0.061\pm 0.015\, ( 0.061^{+0.029}_{-0.029} ) $ & $  0.060\pm 0.013\, ( 0.060^{+0.026}_{-0.026} ) $ \\ 
$ n_\mathrm{s}  $ & $  0.9590\pm 0.0084\, ( 0.959^{+0.017}_{-0.016} ) $ & $  0.9662\pm 0.0066\, ( 0.966^{+0.013}_{-0.013} ) $ & $  0.955\pm 0.023\, ( 0.955^{+0.045}_{-0.046} ) $ & $  0.981\pm 0.018\, ( 0.981^{+0.036}_{-0.035} ) $ & $  0.997\pm 0.037\, ( 0.997^{+0.080}_{-0.076} ) $ & $  0.991\pm 0.025\, ( 0.991^{+0.051}_{-0.048} ) $ \\ 
$ \log(10^{10} A_\mathrm{s})  $ & $  3.036\pm 0.017\, ( 3.036^{+0.034}_{-0.034} ) $ & $  3.049\pm 0.016\, ( 3.049^{+0.031}_{-0.030} ) $ & $  3.024\pm 0.034\, ( 3.024^{+0.065}_{-0.068} ) $ & $  3.066\pm 0.020\, ( 3.066^{+0.038}_{-0.040} ) $ & $  3.056\pm 0.032\, ( 3.056^{+0.063}_{-0.065} ) $ & $  3.063\pm 0.028\, ( 3.063^{+0.054}_{-0.056} ) $ \\ 
$ H_0  $ & $  66.3\pm 1.4\, ( 66.3^{+2.8}_{-2.6} ) $ & $  67.6\pm 1.1\, ( 67.6^{+2.1}_{-2.1} ) $ & $  62.2\pm 2.5\, ( 62.2^{+5.0}_{-4.9} ) $ & $  65.7\pm 1.5\, ( 65.7^{+3.1}_{-3.0} ) $ & $  71.3^{+4.1}_{-4.7}\, ( 71^{+9}_{-8} ) $ & $  69.9\pm 2.0\, ( 69.9^{+4.3}_{-4.2} ) $ \\ 
$ \Omega_\mathrm{m}  $ & $  0.3200\pm 0.0096\, ( 0.320^{+0.019}_{-0.019} ) $ & $  0.3113\pm 0.0059\, ( 0.311^{+0.012}_{-0.011} ) $ & $  0.339\pm 0.019\, ( 0.339^{+0.042}_{-0.040} ) $ & $  0.3143\pm 0.0071\, ( 0.314^{+0.014}_{-0.014} ) $ & $  0.287\pm 0.025\, ( 0.287^{+0.050}_{-0.049} ) $ & $  0.3038\pm 0.0074\, ( 0.304^{+0.015}_{-0.014} ) $ \\ 
$ \sigma_8  $ & $  0.8048\pm 0.0097\, ( 0.805^{+0.019}_{-0.019} ) $ & $  0.8108\pm 0.0094\, ( 0.811^{+0.019}_{-0.018} ) $ & $  0.798\pm 0.018\, ( 0.798^{+0.036}_{-0.035} ) $ & $  0.820\pm 0.014\, ( 0.820^{+0.028}_{-0.027} ) $ & $  0.819\pm 0.028\, ( 0.819^{+0.061}_{-0.058} ) $ & $  0.831\pm 0.022\, ( 0.831^{+0.048}_{-0.046} ) $ \\ 
$ r_\mathrm{drag}  $ & $  148.7\pm 1.9\, ( 148.7^{+3.7}_{-3.7} ) $ & $  147.4\pm 1.7\, ( 147.4^{+3.4}_{-3.3} ) $ & $  155.9\pm 3.8\, ( 155.9^{+7.8}_{-7.3} ) $ & $  151.8\pm 3.1\, ( 151.8^{+6.1}_{-5.8} ) $ & $  144.8\pm 4.8\, ( 144.8^{+9.9}_{-10} ) $ & $  143.8\pm 3.8\, ( 143.8^{+7.1}_{-7.5} ) $ \\ 

\hline \hline
\end{tabular} }
\end{center}
\caption{Mean values and $68\%$ ($95\%$)~CL constraints on the six $\Lambda$CDM parameters as well as on some derived ones ($H_0$, $\Omega_{\rm m}$, $\sigma_8$ and $r_{\rm{drag}}$) within a fiducial cosmology with $N_{\rm{eff}}$ a free parameter. The mean values and errors on the relativistic degrees of freedom at decoupling also presented, for the sake of completeness.}
\label{ tab.results.lcdm_nnu }
\end{table*}

\begin{table*}[h]
\begin{center}
\renewcommand{\arraystretch}{1.5}
\resizebox{\textwidth}{!}{
\begin{tabular}{l c c c c c c c c c c c c c c c }
\hline
\textbf{Parameter} & \textbf{ Planck } & \textbf{ Planck+low-z } & \textbf{ ACT } & \textbf{ ACT+lowz } & \textbf{ SPT } & \textbf{ SPT+low-z } \\ 
\hline\hline

$ \alpha_s  $ & $  -0.0049\pm 0.0067\, ( -0.005^{+0.013}_{-0.013} ) $ & $  -0.0046\pm 0.0068\, ( -0.005^{+0.013}_{-0.013} ) $ & $  0.066\pm 0.023\, ( 0.066^{+0.045}_{-0.046} ) $ & $  0.065\pm 0.024\, ( 0.065^{+0.046}_{-0.046} ) $ & $  -0.071\pm 0.041\, ( -0.071^{+0.079}_{-0.081} ) $ & $  -0.070\pm 0.040\, ( -0.070^{+0.079}_{-0.079} ) $ \\ 
$ \Omega_\mathrm{b} h^2  $ & $  0.02241\pm 0.00015\, ( 0.02241^{+0.00030}_{-0.00029} ) $ & $  0.02246\pm 0.00014\, ( 0.02246^{+0.00027}_{-0.00028} ) $ & $  0.02134\pm 0.00032\, ( 0.02134^{+0.00061}_{-0.00062} ) $ & $  0.02133\pm 0.00032\, ( 0.02133^{+0.00063}_{-0.00061} ) $ & $  0.02229\pm 0.00031\, ( 0.02229^{+0.00062}_{-0.00061} ) $ & $  0.02228\pm 0.00031\, ( 0.02228^{+0.00062}_{-0.00062} ) $ \\ 
$ \Omega_\mathrm{c} h^2  $ & $  0.1200\pm 0.0012\, ( 0.1200^{+0.0024}_{-0.0023} ) $ & $  0.11935\pm 0.00087\, ( 0.1194^{+0.0017}_{-0.0017} ) $ & $  0.1188\pm 0.0021\, ( 0.1188^{+0.0043}_{-0.0041} ) $ & $  0.1189\pm 0.0012\, ( 0.1189^{+0.0023}_{-0.0023} ) $ & $  0.1175\pm 0.0040\, ( 0.1175^{+0.0081}_{-0.0078} ) $ & $  0.1179\pm 0.0013\, ( 0.1179^{+0.0026}_{-0.0026} ) $ \\ 
$ 100\theta_\mathrm{MC}  $ & $  1.04090\pm 0.00031\, ( 1.04090^{+0.00060}_{-0.00061} ) $ & $  1.04100\pm 0.00029\, ( 1.04100^{+0.00057}_{-0.00055} ) $ & $  1.04225\pm 0.00067\, ( 1.0422^{+0.0013}_{-0.0013} ) $ & $  1.04222\pm 0.00063\, ( 1.0422^{+0.0013}_{-0.0012} ) $ & $  1.04010\pm 0.00077\, ( 1.0401^{+0.0015}_{-0.0015} ) $ & $  1.04014\pm 0.00068\, ( 1.0401^{+0.0013}_{-0.0014} ) $ \\ 
$ \tau_\mathrm{reio}  $ & $  0.0553\pm 0.0077\, ( 0.055^{+0.016}_{-0.015} ) $ & $  0.0586\pm 0.0074\, ( 0.059^{+0.015}_{-0.014} ) $ & $  0.059\pm 0.015\, ( 0.059^{+0.029}_{-0.028} ) $ & $  0.061\pm 0.011\, ( 0.061^{+0.022}_{-0.022} ) $ & $  0.062\pm 0.015\, ( 0.062^{+0.029}_{-0.029} ) $ & $  0.068\pm 0.014\, ( 0.068^{+0.027}_{-0.027} ) $ \\ 
$ n_\mathrm{s}  $ & $  0.9641\pm 0.0043\, ( 0.9641^{+0.0086}_{-0.0085} ) $ & $  0.9659\pm 0.0038\, ( 0.9659^{+0.0074}_{-0.0076} ) $ & $  0.973\pm 0.014\, ( 0.973^{+0.028}_{-0.027} ) $ & $  0.974\pm 0.014\, ( 0.974^{+0.028}_{-0.027} ) $ & $  1.006\pm 0.026\, ( 1.006^{+0.051}_{-0.051} ) $ & $  1.005\pm 0.026\, ( 1.005^{+0.050}_{-0.050} ) $ \\ 
$ \log(10^{10} A_\mathrm{s})  $ & $  3.048\pm 0.015\, ( 3.048^{+0.031}_{-0.029} ) $ & $  3.053\pm 0.015\, ( 3.053^{+0.030}_{-0.028} ) $ & $  3.044\pm 0.027\, ( 3.044^{+0.052}_{-0.053} ) $ & $  3.047\pm 0.022\, ( 3.047^{+0.042}_{-0.042} ) $ & $  3.052\pm 0.031\, ( 3.052^{+0.059}_{-0.060} ) $ & $  3.065\pm 0.029\, ( 3.065^{+0.057}_{-0.056} ) $ \\ 
$ H_0  $ & $  67.35\pm 0.54\, ( 67.4^{+1.1}_{-1.1} ) $ & $  67.67\pm 0.40\, ( 67.67^{+0.77}_{-0.77} ) $ & $  67.36\pm 0.89\, ( 67.4^{+1.7}_{-1.8} ) $ & $  67.33\pm 0.49\, ( 67.33^{+0.96}_{-0.94} ) $ & $  67.9\pm 1.6\, ( 67.9^{+3.2}_{-3.1} ) $ & $  67.76\pm 0.52\, ( 67.8^{+1.0}_{-1.0} ) $ \\ 
$ \Omega_\mathrm{m}  $ & $  0.3156\pm 0.0074\, ( 0.316^{+0.015}_{-0.014} ) $ & $  0.3112\pm 0.0053\, ( 0.311^{+0.010}_{-0.010} ) $ & $  0.311\pm 0.013\, ( 0.311^{+0.026}_{-0.024} ) $ & $  0.3108\pm 0.0065\, ( 0.311^{+0.013}_{-0.012} ) $ & $  0.305\pm 0.023\, ( 0.305^{+0.050}_{-0.048} ) $ & $  0.3068\pm 0.0071\, ( 0.307^{+0.014}_{-0.014} ) $ \\ 
$ \sigma_8  $ & $  0.8112\pm 0.0060\, ( 0.811^{+0.012}_{-0.012} ) $ & $  0.8116\pm 0.0059\, ( 0.812^{+0.012}_{-0.011} ) $ & $  0.8288\pm 0.0086\, ( 0.829^{+0.017}_{-0.017} ) $ & $  0.8302\pm 0.0080\, ( 0.830^{+0.016}_{-0.016} ) $ & $  0.804\pm 0.017\, ( 0.804^{+0.033}_{-0.034} ) $ & $  0.810\pm 0.012\, ( 0.810^{+0.023}_{-0.023} ) $ \\ 
$ r_\mathrm{drag}  $ & $  147.05\pm 0.27\, ( 147.05^{+0.53}_{-0.53} ) $ & $  147.18\pm 0.23\, ( 147.18^{+0.45}_{-0.44} ) $ & $  148.58\pm 0.66\, ( 148.6^{+1.3}_{-1.3} ) $ & $  148.58\pm 0.49\, ( 148.58^{+0.98}_{-0.96} ) $ & $  147.9\pm 1.1\, ( 147.9^{+2.2}_{-2.2} ) $ & $  147.77\pm 0.52\, ( 147.8^{+1.0}_{-1.0} ) $ \\ 

\hline \hline
\end{tabular} }
\end{center}
\caption{Mean values and $68\%$ ($95\%$)~CL constraints on the six $\Lambda$CDM parameters as well as on some derived ones ($H_0$, $\Omega_{\rm m}$, $\sigma_8$ and $r_{\rm{drag}}$) within a fiducial cosmology with a running of the scalar spectral index. The mean values and errors on $\alpha_{\rm s}$ are also presented, for the sake of completeness.}
\label{ tab.results.lcdm_nrun }
\end{table*}

\begin{table*}[h]
\begin{center}
\renewcommand{\arraystretch}{1.5}
\resizebox{\textwidth}{!}{
\begin{tabular}{l c c c c c c c c c c c c c c c }
\hline
\textbf{Parameter} & \textbf{ Planck } & \textbf{ Planck+low-z } & \textbf{ ACT } & \textbf{ ACT+lowz } & \textbf{ SPT } & \textbf{ SPT+low-z } \\ 
\hline\hline

$ \Omega_k  $ & $  -0.0104\pm 0.0065\, ( -0.010^{+0.014}_{-0.014} ) $ & $  0.0006\pm 0.0017\, ( 0.0006^{+0.0033}_{-0.0035} ) $ & $  -0.010^{+0.017}_{-0.015}\, ( -0.010^{+0.031}_{-0.033} ) $ & $  0.0000\pm 0.0029\, ( 0.0000^{+0.0057}_{-0.0056} ) $ & $  0.020^{+0.015}_{-0.012}\, ( 0.020^{+0.027}_{-0.029} ) $ & $  0.0018\pm 0.0034\, ( 0.0018^{+0.0068}_{-0.0065} ) $ \\ 
$ \Omega_\mathrm{b} h^2  $ & $  0.02249\pm 0.00016\, ( 0.02249^{+0.00031}_{-0.00031} ) $ & $  0.02240\pm 0.00015\, ( 0.02240^{+0.00029}_{-0.00029} ) $ & $  0.02165\pm 0.00030\, ( 0.02165^{+0.00061}_{-0.00058} ) $ & $  0.02162\pm 0.00029\, ( 0.02162^{+0.00058}_{-0.00057} ) $ & $  0.02212\pm 0.00032\, ( 0.02212^{+0.00063}_{-0.00062} ) $ & $  0.02222\pm 0.00031\, ( 0.02222^{+0.00060}_{-0.00061} ) $ \\ 
$ \Omega_\mathrm{c} h^2  $ & $  0.1185\pm 0.0015\, ( 0.1185^{+0.0029}_{-0.0028} ) $ & $  0.1196\pm 0.0013\, ( 0.1196^{+0.0025}_{-0.0025} ) $ & $  0.1167\pm 0.0047\, ( 0.1167^{+0.0092}_{-0.0089} ) $ & $  0.1191\pm 0.0027\, ( 0.1191^{+0.0054}_{-0.0051} ) $ & $  0.1220\pm 0.0056\, ( 0.122^{+0.011}_{-0.011} ) $ & $  0.1195\pm 0.0035\, ( 0.1195^{+0.0070}_{-0.0068} ) $ \\ 
$ 100\theta_\mathrm{MC}  $ & $  1.04107\pm 0.00032\, ( 1.04107^{+0.00062}_{-0.00062} ) $ & $  1.04095\pm 0.00031\, ( 1.04095^{+0.00062}_{-0.00061} ) $ & $  1.04231\pm 0.00075\, ( 1.0423^{+0.0015}_{-0.0015} ) $ & $  1.04213\pm 0.00069\, ( 1.0421^{+0.0013}_{-0.0014} ) $ & $  1.03983\pm 0.00080\, ( 1.0398^{+0.0016}_{-0.0016} ) $ & $  1.04004\pm 0.00075\, ( 1.0400^{+0.0015}_{-0.0015} ) $ \\ 
$ \tau_\mathrm{reio}  $ & $  0.0493\pm 0.0084\, ( 0.049^{+0.016}_{-0.017} ) $ & $  0.0568\pm 0.0071\, ( 0.057^{+0.015}_{-0.013} ) $ & $  0.065\pm 0.015\, ( 0.065^{+0.030}_{-0.029} ) $ & $  0.070\pm 0.012\, ( 0.070^{+0.024}_{-0.024} ) $ & $  0.063\pm 0.015\, ( 0.063^{+0.029}_{-0.029} ) $ & $  0.061\pm 0.014\, ( 0.061^{+0.028}_{-0.027} ) $ \\ 
$ n_\mathrm{s}  $ & $  0.9688\pm 0.0046\, ( 0.9688^{+0.0089}_{-0.0090} ) $ & $  0.9661\pm 0.0043\, ( 0.9661^{+0.0084}_{-0.0084} ) $ & $  1.004\pm 0.016\, ( 1.004^{+0.032}_{-0.031} ) $ & $  0.997\pm 0.012\, ( 0.997^{+0.024}_{-0.024} ) $ & $  0.958\pm 0.019\, ( 0.958^{+0.038}_{-0.037} ) $ & $  0.966\pm 0.017\, ( 0.966^{+0.033}_{-0.032} ) $ \\ 
$ \log(10^{10} A_\mathrm{s})  $ & $  3.030\pm 0.017\, ( 3.030^{+0.034}_{-0.036} ) $ & $  3.049\pm 0.014\, ( 3.049^{+0.028}_{-0.027} ) $ & $  3.053\pm 0.034\, ( 3.053^{+0.067}_{-0.066} ) $ & $  3.072\pm 0.021\, ( 3.072^{+0.040}_{-0.041} ) $ & $  3.072\pm 0.035\, ( 3.072^{+0.069}_{-0.069} ) $ & $  3.058\pm 0.029\, ( 3.058^{+0.057}_{-0.055} ) $ \\ 
$ H_0  $ & $  63.6\pm 2.2\, ( 63.6^{+4.6}_{-4.3} ) $ & $  67.82\pm 0.57\, ( 67.8^{+1.1}_{-1.1} ) $ & $  64.5\pm 4.5\, ( 65^{+9}_{-8} ) $ & $  67.44\pm 0.65\, ( 67.4^{+1.3}_{-1.2} ) $ & $  81^{+10}_{-9}\, ( 81^{+20}_{-20} ) $ & $  67.98\pm 0.64\, ( 68.0^{+1.3}_{-1.3} ) $ \\ 
$ \Omega_\mathrm{m}  $ & $  0.351\pm 0.024\, ( 0.351^{+0.048}_{-0.045} ) $ & $  0.3103\pm 0.0056\, ( 0.310^{+0.011}_{-0.011} ) $ & $  0.338\pm 0.040\, ( 0.338^{+0.087}_{-0.083} ) $ & $  0.3109\pm 0.0067\, ( 0.311^{+0.013}_{-0.013} ) $ & $  0.229^{+0.047}_{-0.055}\, ( 0.229^{+0.11}_{-0.098} ) $ & $  0.3081\pm 0.0079\, ( 0.308^{+0.016}_{-0.015} ) $ \\ 
$ \sigma_8  $ & $  0.795\pm 0.011\, ( 0.795^{+0.022}_{-0.023} ) $ & $  0.8125\pm 0.0068\, ( 0.813^{+0.013}_{-0.013} ) $ & $  0.814\pm 0.030\, ( 0.814^{+0.057}_{-0.057} ) $ & $  0.8340\pm 0.0094\, ( 0.834^{+0.019}_{-0.019} ) $ & $  0.836\pm 0.030\, ( 0.836^{+0.058}_{-0.059} ) $ & $  0.816\pm 0.016\, ( 0.816^{+0.031}_{-0.031} ) $ \\ 
$ r_\mathrm{drag}  $ & $  147.36\pm 0.30\, ( 147.36^{+0.59}_{-0.60} ) $ & $  147.16\pm 0.28\, ( 147.16^{+0.56}_{-0.54} ) $ & $  148.8\pm 1.3\, ( 148.8^{+2.5}_{-2.4} ) $ & $  148.19\pm 0.76\, ( 148.2^{+1.5}_{-1.5} ) $ & $  146.9\pm 1.4\, ( 146.9^{+2.8}_{-2.8} ) $ & $  147.42\pm 0.98\, ( 147.4^{+1.9}_{-1.9} ) $ \\ 

\hline \hline
\end{tabular} }
\end{center}
\caption{Mean values and $68\%$ ($95\%$)~CL constraints on the six $\Lambda$CDM parameters as well as on some derived ones ($H_0$, $\Omega_{\rm m}$, $\sigma_8$ and $r_{\rm{drag}}$) within a fiducial cosmology with a non-zero spatial curveture. The mean values and errors on the curvature parameter $\Omega_{\rm k}$ are also presented, for the sake of completeness.}
\label{ tab.results.kcdm }
\end{table*}

\begin{table*}[h]
\begin{center}
\renewcommand{\arraystretch}{1.5}
\resizebox{\textwidth}{!}{
\begin{tabular}{l c c c c c c c c c c c c c c c }
\hline
\textbf{Parameter} & \textbf{ Planck } & \textbf{ Planck+low-z } & \textbf{ ACT } & \textbf{ ACT+lowz } & \textbf{ SPT } & \textbf{ SPT+low-z } \\ 
\hline\hline

$ A_{\rm lens}  $ & $  1.071\pm 0.041\, ( 1.071^{+0.084}_{-0.078} ) $ & $  1.061\pm 0.035\, ( 1.061^{+0.070}_{-0.066} ) $ & $  1.081\pm 0.092\, ( 1.08^{+0.20}_{-0.19} ) $ & $  1.020\pm 0.045\, ( 1.020^{+0.092}_{-0.085} ) $ & $  0.85\pm 0.10\, ( 0.85^{+0.22}_{-0.21} ) $ & $  0.885\pm 0.077\, ( 0.88^{+0.16}_{-0.15} ) $ \\ 
$ \Omega_\mathrm{b} h^2  $ & $  0.02251\pm 0.00017\, ( 0.02251^{+0.00033}_{-0.00032} ) $ & $  0.02250\pm 0.00014\, ( 0.02250^{+0.00027}_{-0.00027} ) $ & $  0.02164\pm 0.00030\, ( 0.02164^{+0.00059}_{-0.00056} ) $ & $  0.02161\pm 0.00029\, ( 0.02161^{+0.00057}_{-0.00056} ) $ & $  0.02213\pm 0.00032\, ( 0.02213^{+0.00064}_{-0.00063} ) $ & $  0.02218\pm 0.00031\, ( 0.02218^{+0.00063}_{-0.00060} ) $ \\ 
$ \Omega_\mathrm{c} h^2  $ & $  0.1182\pm 0.0015\, ( 0.1182^{+0.0030}_{-0.0030} ) $ & $  0.11851\pm 0.00097\, ( 0.1185^{+0.0019}_{-0.0019} ) $ & $  0.1160\pm 0.0045\, ( 0.1160^{+0.0089}_{-0.0084} ) $ & $  0.1188\pm 0.0013\, ( 0.1188^{+0.0026}_{-0.0025} ) $ & $  0.1222\pm 0.0059\, ( 0.122^{+0.012}_{-0.011} ) $ & $  0.1182\pm 0.0014\, ( 0.1182^{+0.0027}_{-0.0027} ) $ \\ 
$ 100\theta_\mathrm{MC}  $ & $  1.04109\pm 0.00032\, ( 1.04109^{+0.00063}_{-0.00064} ) $ & $  1.04108\pm 0.00029\, ( 1.04108^{+0.00057}_{-0.00057} ) $ & $  1.04237\pm 0.00074\, ( 1.0424^{+0.0014}_{-0.0014} ) $ & $  1.04212\pm 0.00062\, ( 1.0421^{+0.0012}_{-0.0012} ) $ & $  1.03983\pm 0.00081\, ( 1.0398^{+0.0016}_{-0.0016} ) $ & $  1.04019\pm 0.00067\, ( 1.0402^{+0.0013}_{-0.0013} ) $ \\ 
$ \tau_\mathrm{reio}  $ & $  0.0491\pm 0.0084\, ( 0.049^{+0.018}_{-0.019} ) $ & $  0.0513\pm 0.0080\, ( 0.051^{+0.016}_{-0.016} ) $ & $  0.064\pm 0.015\, ( 0.064^{+0.029}_{-0.030} ) $ & $  0.067\pm 0.014\, ( 0.067^{+0.027}_{-0.028} ) $ & $  0.065\pm 0.015\, ( 0.065^{+0.029}_{-0.029} ) $ & $  0.070\pm 0.014\, ( 0.070^{+0.027}_{-0.028} ) $ \\ 
$ n_\mathrm{s}  $ & $  0.9696\pm 0.0048\, ( 0.9696^{+0.0096}_{-0.0094} ) $ & $  0.9690\pm 0.0038\, ( 0.9690^{+0.0075}_{-0.0074} ) $ & $  1.006\pm 0.016\, ( 1.006^{+0.032}_{-0.032} ) $ & $  0.999\pm 0.013\, ( 0.999^{+0.025}_{-0.024} ) $ & $  0.957\pm 0.020\, ( 0.957^{+0.039}_{-0.038} ) $ & $  0.966\pm 0.015\, ( 0.966^{+0.030}_{-0.030} ) $ \\ 
$ \log(10^{10} A_\mathrm{s})  $ & $  3.028\pm 0.018\, ( 3.028^{+0.038}_{-0.039} ) $ & $  3.034\pm 0.017\, ( 3.034^{+0.033}_{-0.034} ) $ & $  3.048\pm 0.034\, ( 3.048^{+0.066}_{-0.066} ) $ & $  3.063\pm 0.031\, ( 3.063^{+0.060}_{-0.060} ) $ & $  3.076\pm 0.036\, ( 3.076^{+0.070}_{-0.071} ) $ & $  3.076\pm 0.031\, ( 3.076^{+0.062}_{-0.061} ) $ \\ 
$ H_0  $ & $  68.14\pm 0.70\, ( 68.1^{+1.4}_{-1.4} ) $ & $  68.03\pm 0.44\, ( 68.03^{+0.88}_{-0.85} ) $ & $  68.7\pm 1.8\, ( 68.7^{+3.6}_{-3.5} ) $ & $  67.54\pm 0.51\, ( 67.5^{+1.0}_{-1.0} ) $ & $  66.1\pm 2.3\, ( 66.1^{+4.6}_{-4.3} ) $ & $  67.58\pm 0.53\, ( 67.6^{+1.1}_{-1.0} ) $ \\ 
$ \Omega_\mathrm{m}  $ & $  0.3047\pm 0.0092\, ( 0.305^{+0.018}_{-0.018} ) $ & $  0.3062\pm 0.0058\, ( 0.306^{+0.011}_{-0.011} ) $ & $  0.294\pm 0.025\, ( 0.294^{+0.054}_{-0.052} ) $ & $  0.3094\pm 0.0071\, ( 0.309^{+0.014}_{-0.014} ) $ & $  0.334\pm 0.036\, ( 0.334^{+0.079}_{-0.076} ) $ & $  0.3089\pm 0.0074\, ( 0.309^{+0.015}_{-0.014} ) $ \\ 
$ \sigma_8  $ & $  0.7998\pm 0.0088\, ( 0.800^{+0.017}_{-0.018} ) $ & $  0.8029\pm 0.0076\, ( 0.803^{+0.015}_{-0.015} ) $ & $  0.816\pm 0.021\, ( 0.816^{+0.041}_{-0.041} ) $ & $  0.830\pm 0.013\, ( 0.830^{+0.025}_{-0.025} ) $ & $  0.828\pm 0.025\, ( 0.828^{+0.048}_{-0.050} ) $ & $  0.819\pm 0.013\, ( 0.819^{+0.026}_{-0.026} ) $ \\ 
$ r_\mathrm{drag}  $ & $  147.41\pm 0.31\, ( 147.41^{+0.62}_{-0.61} ) $ & $  147.36\pm 0.23\, ( 147.36^{+0.46}_{-0.47} ) $ & $  149.0\pm 1.2\, ( 149.0^{+2.3}_{-2.4} ) $ & $  148.27\pm 0.51\, ( 148.3^{+1.0}_{-0.97} ) $ & $  146.8\pm 1.5\, ( 146.8^{+3.0}_{-2.9} ) $ & $  147.80\pm 0.52\, ( 147.8^{+1.0}_{-1.0} ) $ \\ 

\hline \hline
\end{tabular} }
\end{center}
\caption{Mean values and $68\%$~ ($95\%$) CL constraints on the six $\Lambda$CDM parameters as well as on some derived ones ($H_0$, $\Omega_{\rm m}$, $\sigma_8$ and $r_{\rm{drag}}$) within a fiducial cosmology with a varying lensing amplitude. The mean values and errors on $A_{\rm{lens}}$ are also presented, for the sake of completeness.}
\label{ tab.results.Alens }
\end{table*}

\begin{table*}[h]
\begin{center}
\renewcommand{\arraystretch}{1.5}
\resizebox{\textwidth}{!}{
\begin{tabular}{l c c c c c c c c c c c c c c c }
\hline
\textbf{Parameter} & \textbf{ Planck } & \textbf{ Planck+low-z } & \textbf{ ACT } & \textbf{ ACT+lowz } & \textbf{ SPT } & \textbf{ SPT+low-z } \\ 
\hline\hline

$ w_0  $ & $  -1.55^{+0.26}_{-0.31}\, ( -1.55^{+0.57}_{-0.54} ) $ & $  -0.995\pm 0.024\, ( -0.995^{+0.047}_{-0.048} ) $ & $  -1.43^{+0.35}_{-0.42}\, ( -1.43^{+0.77}_{-0.72} ) $ & $  -0.975\pm 0.027\, ( -0.975^{+0.052}_{-0.055} ) $ & $  -0.76^{+0.58}_{-0.43}\, ( -0.76^{+0.94}_{-1.0} ) $ & $  -0.966\pm 0.029\, ( -0.966^{+0.056}_{-0.057} ) $ \\ 
$ \Omega_\mathrm{b} h^2  $ & $  0.02243\pm 0.00015\, ( 0.02243^{+0.00029}_{-0.00029} ) $ & $  0.02244\pm 0.00013\, ( 0.02244^{+0.00027}_{-0.00026} ) $ & $  0.02160\pm 0.00030\, ( 0.02160^{+0.00058}_{-0.00058} ) $ & $  0.02164\pm 0.00030\, ( 0.02164^{+0.00060}_{-0.00058} ) $ & $  0.02219\pm 0.00032\, ( 0.02219^{+0.00063}_{-0.00062} ) $ & $  0.02225\pm 0.00031\, ( 0.02225^{+0.00062}_{-0.00061} ) $ \\ 
$ \Omega_\mathrm{c} h^2  $ & $  0.1193\pm 0.0012\, ( 0.1193^{+0.0024}_{-0.0024} ) $ & $  0.11921\pm 0.00098\, ( 0.1192^{+0.0019}_{-0.0019} ) $ & $  0.1187\pm 0.0023\, ( 0.1187^{+0.0050}_{-0.0049} ) $ & $  0.1183\pm 0.0015\, ( 0.1183^{+0.0029}_{-0.0029} ) $ & $  0.1194^{+0.0055}_{-0.0062}\, ( 0.119^{+0.012}_{-0.011} ) $ & $  0.1164\pm 0.0018\, ( 0.1164^{+0.0035}_{-0.0035} ) $ \\ 
$ 100\theta_\mathrm{MC}  $ & $  1.04098\pm 0.00031\, ( 1.04098^{+0.00060}_{-0.00060} ) $ & $  1.04102\pm 0.00029\, ( 1.04102^{+0.00057}_{-0.00058} ) $ & $  1.04213\pm 0.00066\, ( 1.0421^{+0.0013}_{-0.0013} ) $ & $  1.04222\pm 0.00064\, ( 1.0422^{+0.0012}_{-0.0012} ) $ & $  1.04004\pm 0.00080\, ( 1.0400^{+0.0015}_{-0.0016} ) $ & $  1.04033\pm 0.00069\, ( 1.0403^{+0.0013}_{-0.0014} ) $ \\ 
$ \tau_\mathrm{reio}  $ & $  0.0524\pm 0.0074\, ( 0.052^{+0.015}_{-0.015} ) $ & $  0.0579\pm 0.0074\, ( 0.058^{+0.015}_{-0.014} ) $ & $  0.066\pm 0.014\, ( 0.066^{+0.027}_{-0.027} ) $ & $  0.076\pm 0.012\, ( 0.076^{+0.025}_{-0.024} ) $ & $  0.061\pm 0.015\, ( 0.061^{+0.029}_{-0.029} ) $ & $  0.067\pm 0.014\, ( 0.067^{+0.027}_{-0.027} ) $ \\ 
$ n_\mathrm{s}  $ & $  0.9666\pm 0.0042\, ( 0.9666^{+0.0082}_{-0.0081} ) $ & $  0.9671\pm 0.0038\, ( 0.9671^{+0.0074}_{-0.0073} ) $ & $  0.9997\pm 0.012\, ( 0.9997^{+0.024}_{-0.024} ) $ & $  0.998\pm 0.012\, ( 0.998^{+0.023}_{-0.023} ) $ & $  0.964\pm 0.019\, ( 0.964^{+0.036}_{-0.039} ) $ & $  0.972\pm 0.015\, ( 0.972^{+0.030}_{-0.030} ) $ \\ 
$ \log(10^{10} A_\mathrm{s})  $ & $  3.038\pm 0.014\, ( 3.038^{+0.029}_{-0.028} ) $ & $  3.051\pm 0.014\, ( 3.051^{+0.029}_{-0.027} ) $ & $  3.060\pm 0.026\, ( 3.060^{+0.051}_{-0.051} ) $ & $  3.082\pm 0.023\, ( 3.082^{+0.045}_{-0.044} ) $ & $  3.060\pm 0.035\, ( 3.060^{+0.072}_{-0.066} ) $ & $  3.063\pm 0.029\, ( 3.063^{+0.057}_{-0.057} ) $ \\ 
$ H_0  $ & $ > 81.8\, (> 69.7 ) $ & $  67.54\pm 0.65\, ( 67.5^{+1.3}_{-1.3} ) $ & $ > 76.4\, (> 60.7 ) $ & $  67.02\pm 0.68\, ( 67.0^{+1.4}_{-1.3} ) $ & $  62^{+10}_{-20}\, ( 62^{+30}_{-30} ) $ & $  67.27\pm 0.67\, ( 67.3^{+1.3}_{-1.3} ) $ \\ 
$ \Omega_\mathrm{m}  $ & $  0.200^{+0.043}_{-0.061}\, ( 0.20^{+0.11}_{-0.10} ) $ & $  0.3120\pm 0.0065\, ( 0.312^{+0.013}_{-0.012} ) $ & $  0.225^{+0.066}_{-0.10}\, ( 0.23^{+0.17}_{-0.17} ) $ & $  0.3131\pm 0.0071\, ( 0.313^{+0.015}_{-0.013} ) $ & $  0.43\pm 0.17\, ( 0.43^{+0.35}_{-0.36} ) $ & $  0.3079\pm 0.0072\, ( 0.308^{+0.014}_{-0.014} ) $ \\ 
$ \sigma_8  $ & $  0.961^{+0.085}_{-0.072}\, ( 0.96^{+0.15}_{-0.16} ) $ & $  0.8099\pm 0.0090\, ( 0.810^{+0.018}_{-0.018} ) $ & $  0.95^{+0.12}_{-0.10}\, ( 0.95^{+0.20}_{-0.22} ) $ & $  0.828\pm 0.011\, ( 0.828^{+0.021}_{-0.021} ) $ & $  0.74^{+0.12}_{-0.16}\, ( 0.74^{+0.28}_{-0.26} ) $ & $  0.798\pm 0.016\, ( 0.798^{+0.031}_{-0.031} ) $ \\
$ r_\mathrm{drag}  $ & $  147.22\pm 0.27\, ( 147.22^{+0.52}_{-0.52} ) $ & $  147.24\pm 0.24\, ( 147.24^{+0.47}_{-0.46} ) $ & $  148.32\pm 0.68\, ( 148.3^{+1.3}_{-1.4} ) $ & $  148.38\pm 0.51\, ( 148.38^{+0.98}_{-1.0} ) $ & $  147.5\pm 1.4\, ( 147.5^{+3.0}_{-3.1} ) $ & $  148.21\pm 0.59\, ( 148.2^{+1.2}_{-1.1} ) $ \\ 

\hline \hline
\end{tabular} }
\end{center}
\caption{Mean values and $68\%$ ($95\%$) CL constraints on the six $\Lambda$CDM parameters as well as on some derived ones ($H_0$, $\Omega_{\rm m}$, $\sigma_8$ and $r_{\rm{drag}}$) within a fiducial cosmology with a varying dark energy equation of state.  The mean values and errors on $w$ are also presented, for the sake of completeness.}
\label{ tab.results.w }
\end{table*}

\begin{table*}[h]
\begin{center}
\renewcommand{\arraystretch}{1.5}
\resizebox{\textwidth}{!}{
\begin{tabular}{l c c c c c c c c c c c c c c c }
\hline
\textbf{Parameter} & \textbf{ Planck } & \textbf{ Planck+low-z } & \textbf{ ACT } & \textbf{ ACT+lowz } & \textbf{ SPT } & \textbf{ SPT+low-z } \\ 
\hline\hline

$ w_0  $ & $  -1.24\pm 0.50\, ( -1.2^{+1.1}_{-1.0} ) $ & $  -0.859\pm 0.061\, ( -0.86^{+0.12}_{-0.12} ) $ & $  -1.13^{+0.62}_{-0.69}\, ( -1.1^{+1.3}_{-1.2} ) $ & $  -0.880\pm 0.062\, ( -0.88^{+0.12}_{-0.12} ) $ & $  -0.54^{+0.93}_{-0.67}\, ( -0.5^{+1.5}_{-1.4} ) $ & $  -0.878\pm 0.064\, ( -0.88^{+0.13}_{-0.12} ) $ \\ 
$ w_{a}  $ & $ < 1.21 $ & $  -0.58^{+0.26}_{-0.23}\, ( -0.58^{+0.46}_{-0.50} ) $ & $ < -0.304\, (--) $ & $  -0.44^{+0.30}_{-0.24}\, ( -0.44^{+0.57}_{-0.59} ) $ & $  -0.8^{+1.2}_{-1.5}\, (--) $ & $  -0.45^{+0.31}_{-0.26}\, (-0.45^{+0.61}_{-0.63}) $ \\ 
$ \Omega_\mathrm{b} h^2  $ & $  0.02244\pm 0.00015\, ( 0.02244^{+0.00029}_{-0.00028} ) $ & $  0.02238\pm 0.00014\, ( 0.02238^{+0.00027}_{-0.00027} ) $ & $  0.02160\pm 0.00030\, ( 0.02160^{+0.00060}_{-0.00058} ) $ & $  0.02163\pm 0.00029\, ( 0.02163^{+0.00057}_{-0.00057} ) $ & $  0.02221\pm 0.00032\, ( 0.02221^{+0.00062}_{-0.00063} ) $ & $  0.02223\pm 0.00031\, ( 0.02223^{+0.00061}_{-0.00061} ) $ \\ 
$ \Omega_\mathrm{c} h^2  $ & $  0.1192\pm 0.0012\, ( 0.1192^{+0.0024}_{-0.0024} ) $ & $  0.1199\pm 0.0010\, ( 0.1199^{+0.0020}_{-0.0020} ) $ & $  0.1187\pm 0.0024\, ( 0.1187^{+0.0053}_{-0.0052} ) $ & $  0.1195\pm 0.0016\, ( 0.1195^{+0.0032}_{-0.0032} ) $ & $  0.1182\pm 0.0049\, ( 0.118^{+0.011}_{-0.010} ) $ & $  0.1184\pm 0.0020\, ( 0.1184^{+0.0039}_{-0.0041} ) $ \\ 
$ 100\theta_\mathrm{MC}  $ & $  1.04100\pm 0.00031\, ( 1.04100^{+0.00061}_{-0.00062} ) $ & $  1.04092\pm 0.00030\, ( 1.04092^{+0.00058}_{-0.00057} ) $ & $  1.04214\pm 0.00066\, ( 1.0421^{+0.0013}_{-0.0013} ) $ & $  1.04202\pm 0.00062\, ( 1.0420^{+0.0012}_{-0.0012} ) $ & $  1.04014\pm 0.00078\, ( 1.0401^{+0.0015}_{-0.0016} ) $ & $  1.04010\pm 0.00070\, ( 1.0401^{+0.0014}_{-0.0014} ) $ \\ 
$ \tau_\mathrm{reio}  $ & $  0.0521\pm 0.0075\, ( 0.052^{+0.015}_{-0.015} ) $ & $  0.0536\pm 0.0073\, ( 0.054^{+0.015}_{-0.014} ) $ & $  0.065\pm 0.014\, ( 0.065^{+0.028}_{-0.028} ) $ & $  0.066\pm 0.013\, ( 0.066^{+0.026}_{-0.026} ) $ & $  0.061\pm 0.015\, ( 0.061^{+0.029}_{-0.029} ) $ & $  0.061\pm 0.014\, ( 0.061^{+0.028}_{-0.028} ) $ \\ 
$ n_\mathrm{s}  $ & $  0.9668\pm 0.0041\, ( 0.9668^{+0.0081}_{-0.0079} ) $ & $  0.9652\pm 0.0038\, ( 0.9652^{+0.0075}_{-0.0074} ) $ & $  0.9998\pm 0.012\, ( 0.9998^{+0.024}_{-0.025} ) $ & $  0.997\pm 0.011\, ( 0.997^{+0.022}_{-0.022} ) $ & $  0.967\pm 0.018\, ( 0.967^{+0.035}_{-0.037} ) $ & $  0.968\pm 0.015\, ( 0.968^{+0.030}_{-0.030} ) $ \\ 
$ \log(10^{10} A_\mathrm{s})  $ & $  3.037\pm 0.015\, ( 3.037^{+0.029}_{-0.029} ) $ & $  3.043\pm 0.014\, ( 3.043^{+0.029}_{-0.028} ) $ & $  3.059\pm 0.026\, ( 3.059^{+0.052}_{-0.051} ) $ & $  3.063\pm 0.025\, ( 3.063^{+0.048}_{-0.048} ) $ & $  3.055\pm 0.034\, ( 3.055^{+0.068}_{-0.065} ) $ & $  3.055\pm 0.029\, ( 3.055^{+0.057}_{-0.057} ) $ \\ 
$ H_0  $ & $ > 79.2\, (> 64.1 ) $ & $  67.62\pm 0.64\, ( 67.6^{+1.2}_{-1.3} ) $ & $ > 73.4\, (> 56.1 ) $ & $  67.11\pm 0.66\, ( 67.1^{+1.3}_{-1.3} ) $ & $  63^{+10}_{-20}\, ( 63^{+30}_{-30} ) $ & $  67.35\pm 0.68\, ( 67.4^{+1.3}_{-1.3} ) $ \\ 
$ \Omega_\mathrm{m}  $ & $  0.214^{+0.054}_{-0.084}\, ( 0.21^{+0.14}_{-0.14} ) $ & $  0.3127\pm 0.0064\, ( 0.313^{+0.013}_{-0.012} ) $ & $  0.245^{+0.079}_{-0.13}\, ( 0.25^{+0.21}_{-0.21} ) $ & $  0.3149\pm 0.0071\, ( 0.315^{+0.014}_{-0.014} ) $ & $  0.42\pm 0.18\, ( 0.42^{+0.38}_{-0.38} ) $ & $  0.3115\pm 0.0076\, ( 0.311^{+0.015}_{-0.015} ) $ \\ 
$ \sigma_8  $ & $  0.944^{+0.10}_{-0.085}\, ( 0.94^{+0.18}_{-0.19} ) $ & $  0.8161\pm 0.0092\, ( 0.816^{+0.018}_{-0.018} ) $ & $  0.93^{+0.13}_{-0.11}\, ( 0.93^{+0.23}_{-0.24} ) $ & $  0.833\pm 0.011\, ( 0.833^{+0.021}_{-0.021} ) $ & $  0.74^{+0.13}_{-0.16}\, ( 0.74^{+0.29}_{-0.27} ) $ & $  0.812\pm 0.017\, ( 0.812^{+0.032}_{-0.033} ) $ \\ 
$ r_\mathrm{drag}  $ & $  147.23\pm 0.27\, ( 147.23^{+0.52}_{-0.53} ) $ & $  147.11\pm 0.24\, ( 147.11^{+0.46}_{-0.46} ) $ & $  148.32\pm 0.72\, ( 148.3^{+1.6}_{-1.6} ) $ & $  148.07\pm 0.53\, ( 148.1^{+1.0}_{-1.0} ) $ & $  147.7\pm 1.3\, ( 147.7^{+2.8}_{-2.9} ) $ & $  147.69\pm 0.65\, ( 147.7^{+1.3}_{-1.2} ) $ \\ 

\hline \hline
\end{tabular} }
\end{center}
\caption{Mean values and $68\%$ ($95\%$)~CL constraints on the six $\Lambda$CDM parameters as well as on some derived ones ($H_0$, $\Omega_{\rm m}$, $\sigma_8$ and $r_{\rm{drag}}$) within a fiducial cosmology with a time-varying dark energy equation of state. The mean values and errors on the parameters $w_0$ and $w_{\rm a}$, see Eq.~\ref{eq:cpl}, are also presented, for the sake of completeness.}
\label{ tab.results.w0wa }
\end{table*}

\end{document}